%% file: ms.tex
\documentclass[11pt,preprint]{aastex}
\usepackage{natbib}
\begin{document}

\shorttitle{Millimeter-Wave Census of Taurus Multiples}
\shortauthors{Harris et al.}

\title{A Resolved Census of Millimeter Emission from Taurus Multiple Star Systems}

\author{Robert J. Harris, Sean M. Andrews, David J. Wilner}
\affil{Harvard-Smithsonian Center for Astrophysics, 60 Garden Street, Cambridge, MA 02138}

\and 

\author{Adam L. Kraus\altaffilmark{1}}
\affil{University of Hawaii Institute for Astronomy, 2680 Woodlawn Drive, Honolulu, HI 96822}
\altaffiltext{1}{Hubble Fellow.}

\begin{abstract}
We present a high angular resolution millimeter-wave dust continuum imaging 
survey of circumstellar material associated with the individual components of 
23 multiple star systems in the Taurus-Auriga young cluster.  Combined with 
previous measurements in the literature, these new data permit a comprehensive 
look at how the millimeter luminosity (a rough tracer of disk mass) relates to 
the separation and mass of a stellar companion.  Approximately one third 
(28-37\%) of the individual stars in multiple systems have detectable 
millimeter emission, an incidence rate half that for single stars ($\sim$62\%) 
which does not depend on the number of companions.  There is a strong, positive 
correlation between the luminosity and projected separation ($a_p$) of a 
stellar pair.  Wide pairs ($a_p > 300$\,AU) have a similar luminosity 
distribution as single stars, medium pairs ($a_p \approx 30$-300\,AU) are a 
factor of 5 fainter, and close pairs ($a_p < 30$\,AU) are $\sim$5$\times$ 
fainter yet (aside from a small, but notable population of bright circumbinary 
disks).  In most cases, the emission is dominated by a disk around the primary 
(or a wide tertiary in hierarchical triples), but there is no clear 
relationship between luminosity and stellar mass ratio.  A direct comparison of 
resolved disk sizes with predictions from tidal truncation models yields mixed 
results; some disks are much larger than expected given the projected distances 
of their companions.  We suggest that the presence of a stellar companion 
impacts disk properties at a level comparable to the internal evolution 
mechanisms that operate in an isolated system, with both the multiple star 
formation process itself and star-disk tidal interactions likely playing 
important roles in the evolution of circumstellar material.  From the 
perspective of the mass content of the disk reservoir, we expect that (giant) 
planet formation is inhibited around the components of close pairs or 
secondaries, but should be as likely as for single stars around the primaries 
(or wide tertiaries in hierarchical triples) in more widely-separated multiple 
star systems.  
\end{abstract}
\keywords{binaries --- protoplanetary disks --- stars: formation}

\section{Introduction} 

Many, if not most, stars are born with close companions \citep[][but see \citealt{lada06}]{abt76,dm91,fm92,raghavan10}.  Depending on their orbits, tidal 
interactions between individual stellar components in these multiple systems 
can dominate the evolution of their natal circumstellar material and 
potentially have drastic consequences for the planet formation process 
\citep[e.g.,][]{papaloizou77,lin79a,lin79b,al94}.  But even in these hazardous 
dynamical environments, many young multiples harbor long-lived circumstellar 
material \citep[see][]{duchene07} and a growing number of their more mature 
counterparts are being identified as exoplanet hosts 
\citep[e.g.,][]{patience02,raghavan06,desidera07,doyle11}.  Given the 
prevalence of stellar multiplicity, an improved empirical understanding of the 
dynamical interplay between the stars and disks in these systems -- including 
effects like tidal truncation, stripping, and the orbital evolution of 
companions -- is fundamental for the development of a comprehensive model for 
the formation of stars and planetary systems.  Moreover, constraints on these 
dynamical processes in multiple star systems can be used as high mass-ratio 
touchstones for theoretical work on analogous disk-planet interactions, 
particularly the creation of tidal gaps and subsequent planet migration.  

A wealth of theoretical work suggests that the fate of the circumstellar 
material in a young multiple star system is primarily dependent on the 
separation ($a$) and mass ratio ($q$) of the individual components, as well as 
the orbital eccentricity ($e$) \citep[e.g.,][]{al94}.  Systems with eccentric 
orbits have an enhanced likelihood of star-disk tidal interactions.  For a 
given orbit, a near equal-mass companion ($q \sim 1$) should have a more 
destructive impact on disk material than a low-mass companion.  But in most 
cases, the effects of \{$e$, $q$\} on the circumstellar material are secondary 
to the orbital separation.  Systems with large separations ($a \sim$ hundreds 
of AU) should impart little or no dynamical effects on their circumstellar 
material, leaving disks around each stellar component that are similar to those 
around single stars.  Conversely, individual disks in a small-separation ($a 
\sim$ a few AU or less) system will likely not survive.  Instead, these systems 
can host a circum-multiple disk with a dynamically cleared central cavity out 
to a radius comparable to the stellar separation ($\sim$2-3$a$).  However, most 
multiple systems both in the field and in young clusters have intermediate 
separations \citep[$a \sim$ tens of AU;][]{mathieu00}.  The disks in these 
systems may suffer the most dramatic effects of star-disk interactions, 
resulting in their external truncation at a fraction of the component 
separation ($\sim$0.2-0.5$a$), or their complete dispersal.  

Qualitatively, these theoretical predictions find some observational support.  
Statistical analyses of warm gas and dust diagnostics (accretion signatures 
and/or a near-infrared excess) indicate that the presence of a companion with 
separation $\lesssim$40\,AU may significantly hasten disk dispersal near the 
stars \citep[on $\sim$1-10\,AU scales;][]{cieza09,kraus11b}.  This diminished 
frequency of disk signatures for ``close" multiples was first noted at 
(sub)millimeter wavelengths by \citet{jensen94}, and later confirmed in surveys 
of increasing size and sensitivity to dust emission 
\citep{osterloh95,jensen96,aw05}.  Since the continuum emission at such long 
wavelengths is primarily optically thin \citep[e.g.,][]{beckwith90}, the 
systematically lower millimeter-wave luminosities from close multiples compared 
to systems with wider separations or single stars were taken as compelling 
evidence for decreased disk masses due to tidal truncation or disruption.  
However, that evidence is indirect: those observations relied on single-dish 
photometers that do not resolve the individual stellar components nor their 
disks.  A quantitative investigation of the theory of star-disk interactions 
requires observations that can address how disk masses and sizes depend on the 
properties of the stellar system (particularly $a$ and $q$).  With the right 
combination of angular resolution and mass sensitivity, interferometric 
measurements of the optically-thin millimeter continuum emission from the dusty 
disks in these systems are uniquely qualified for that task.  Aside from a 
small collection of systems 
\citep[e.g.,][]{jensen96b,akeson98,jensen03,patience08}, such data are rare.

In this article, we present a Submillimeter Array (SMA) survey of the 
millimeter-wave continuum emission from 23 young multiple star systems in the 
Taurus-Auriga star formation region.  These data represent the most 
comprehensive resolved census of cool dust emission from the disks that reside 
in young multiple systems to date.  The motivation for the survey sample is 
introduced in \S 2, and the observations and data calibration are reviewed in 
\S 3.  A simple modeling analysis of these data is conducted in \S 4, with a 
focus on retrieving luminosities and sizes from individual disks whenever 
possible.  The results of this imaging survey are synthesized with other 
information in the literature in \S 4 to extract a statistically representative 
view of circumstellar material in multiple star systems.  Based on that 
analysis, we attempt to reconcile the observations with theoretical predictions 
from tidal interaction models in \S 5.  Finally, our key conclusions are 
summarized in \S 6.

\section{The Sample}

Multiplicity searches in the Taurus molecular clouds have a long history of 
success with a variety of techniques, ranging from straightforward direct 
imaging \citep{rz93,wg01,correia06,kraus09} and radial velocity monitoring 
\citep[e.g.,][]{mathieu97} to more specialized methods like lunar occultations 
\citep{simon92,simon95,richichi99}, speckle interferometry 
\citep{ghez93,leinert93}, and most recently aperture-mask interferometry 
\citep{kraus11}.  There is now a reasonably complete census of Taurus multiple 
systems that have angular separations $\rho \approx 0.03$-30\arcsec, $K$-band 
contrast ratios of $\le$6 magnitudes ($\le$4 mags for the systems with the 
smallest separations), and primary spectral types between F0 and M4.  Assuming 
a mean distance of 145\,pc \citep{loinard07,torres07,torres09} and a crude 
estimation of stellar masses \citep[see][]{kraus11}, this region of 
multiplicity parameter-space corresponds to projected separations $a_p \approx 
5$-5000\,AU, stellar mass ratios $q \approx 0.1$-1.0 (well into the brown dwarf 
regime), and primary star masses $M_p \sim 0.2$-2\,M$_{\odot}$.  There are 
currently 71 such multiple ``systems" known in Taurus, consisting of 111 
``pairs" of 179 individual stars.  For the sake of clarity, we adopt a simple 
nomenclature in this article such that a ``system" refers to any group of 
associated stars and a ``pair" is meant as any subset of the system that could 
potentially interact dynamically.  A simple binary (2 stars) counts as 1 system 
and 1 pair.  In a higher-order hierarchical multiple like UZ Tau, we consider 
the 4 stars UZ Tau Ea, Eb, Wa, and Wb to comprise 1 system of 3 pairs based on 
their relative projected separations: Ea$-$Eb, Wa$-$Wb, and Eab$-$Wab (i.e., 
this phenomenological scheme implicitly assumes that pairs like Ea$-$Wa are 
unlikely to interact directly: all such pairs are listed in \S 5).  

The selection criteria for our resolved millimeter-wave imaging survey were 
motivated by practical observational limitations and consist of two 
requirements: (1) a composite system flux density of $\ge$20\,mJy at 
880\,$\mu$m, and (2) an angular separation of $\ge$0\farcs3 for at least one 
pair in the system.  The first criterion is a sensitivity restriction that 
would ensure that our observations could firmly detect (3-5\,$\sigma$) two 
equivalent disks around individual stellar components with the typical 
expected RMS sensitivity of $\sim$2-3\,mJy beam$^{-1}$ (see \S 3).  Flux 
density estimates for unresolved systems were compiled from the single-dish 
survey of \citet{aw05}.  If no suitable 880\,$\mu$m flux density was 
available, the 1.3\,mm measurements of \citet{beckwith90} or 
\citet{osterloh95} were scaled up by a conservative factor of $(1.3/0.88)^2 
\approx 2.2$ based on the median emission ratio at those wavelengths 
\citep{aw05,aw07}.  Note that for the standard optically thin and isothermal 
assumptions for converting luminosity into mass, this sensitivity threshold 
corresponds to $\sim$5\,M$_{\oplus}$ of dust (or a total mass of 
$\sim$1.5\,M$_{\rm Jup}$ for a 100:1 gas-to-dust mass ratio).  The second 
criterion is a resolution restriction set by the longest available baselines 
of the SMA ($\sim$0.5\,km) that ensures we would be able to resolve the 
individual stellar components of a given pair.  Only systems where the angular 
separations of all its constituent pairs are $<$0\farcs3 were excluded.  The 
resulting sample includes systems where some pair has a projected separation 
$a_p > 40$\,AU.  

Of the original 71 systems, 29 exceed the 880\,$\mu$m luminosity selection 
threshold.  Of those 29 systems, only 2 fail to meet the resolution criterion 
(IS Tau and DQ Tau; the latter is a spectroscopic binary).  The resulting 27 
systems are comprised of 52 pairs and a total of 77 individual stars.  We 
elected not to observe 4 of those systems with the SMA (FS Tau/Haro 6-5B, 
XZ/HL Tau, FZ/FY Tau, and V807/GH Tau) because their wide-separation pairs were 
already resolved with single-dish telescopes and their remaining pairs were too 
faint or too close to meet our selection criteria.  The systems in our 
880\,$\mu$m flux- and resolution-limited sample are listed in Table 
\ref{table:sample}.  

\begin{figure}[h]
\plotone{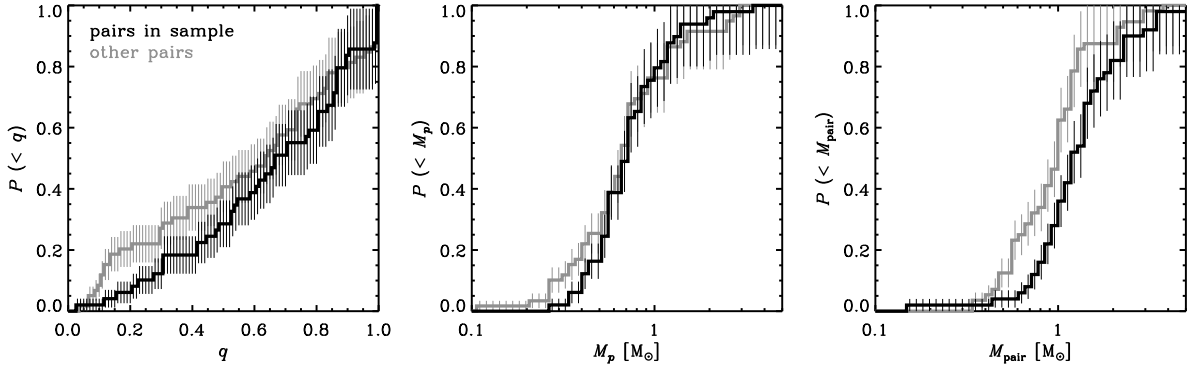}
\figcaption{A comparison of the cumulative distributions of the stellar mass 
ratios ($q$), primary masses ($M_p$), and composite pair masses ($M_{\rm pair} 
= M_p+M_s$) between the systems that were selected ({\it black}) and excluded 
({\it gray}) from our sample.  While our selection criteria do not produce a 
significant bias in $q$ or $M_p$, they tend to include pairs with marginally 
higher {\it total} stellar masses ($M_{\rm pair}$).  \label{fig:biases}}
\end{figure}

Given the practical restrictions that were imposed in its construction, it is 
important to investigate the resulting sample for unintended biases.  Since our 
selection criteria do not specifically address the stellar masses in these 
systems, the key potential biases could be related to the mass ratios of pairs 
($q$), the ``primary" masses ($M_p$; meaning the mass of the brighter component 
of a pair), or the total pair masses (the sum of the primary and secondary 
components of a pair, $M_{\rm pair} = M_p + M_s$).  Figure \ref{fig:biases} 
directly compares the cumulative distributions of $q$, $M_p$, and $M_{\rm 
pair}$ for the pairs included in our sample ({\it black}) against those that 
were excluded due to their low millimeter luminosities and/or small angular 
separations ({\it gray}).  Note that masses were typically determined from the 
\citet{baraffe98} models by \citet{kraus11}; complete references are provided 
in \S 5.  A two-sided Kolmogorov-Smirnov (K-S) test confirms that our sample is 
not biased with respect to $q$ or $M_p$, with probabilities that the sample 
multiples and other multiples are drawn from the same parent distributions of 
mass ratios or primary masses being $\sim$73\%\ and 20\%, respectively.  
However, the distribution of $M_{\rm pair}$ for the pairs in our sample are 
found to be drawn from a marginally different -- and systematically {\it 
higher} -- parent distribution than their counterparts that were not selected, 
with a K-S probability of only 2\%\ (although it is worthwhile to keep in mind 
that these individual stellar mass estimates are crude).

\section{Observations and Data Reduction}

The 23 multiple systems listed in Table \ref{table:sample} (not including the 
additional 4 systems in italics; see \S 2) were observed with the SMA 
interferometer \citep{ho04} in a variety of observing configurations and 
receiver settings over the past 6 years, with most of the data obtained in the 
past 20 months.  An observing log is provided in Table \ref{table:log}.  All 
systems were observed in the compact (C) array configuration, with baseline 
lengths of 8-70\,m.  Additional measurements were made with the extended (E: 
28-226\,m baselines) or very extended (V: 68-509\,m baselines) configurations 
for the systems that contain pairs with smaller angular separations.  Some of 
these data were presented in previous work \citep{aw07b,andrews11}.  All but 4 
of these 23 systems were observed with the 345\,GHz receivers: the pairs in the 
HP Tau, GI/GK Tau, MHO 1/2, and GG Tau systems that we aimed to probe have wide 
enough separations that they were instead observed with slightly lower 
resolution at 230\,GHz.  We made some additional observations of systems that 
were {\it not} in our sample, since they had not yet been observed at 
millimeter wavelengths (see \S 5).   

\begin{figure}[ht!]
\epsscale{0.9}
\plotone{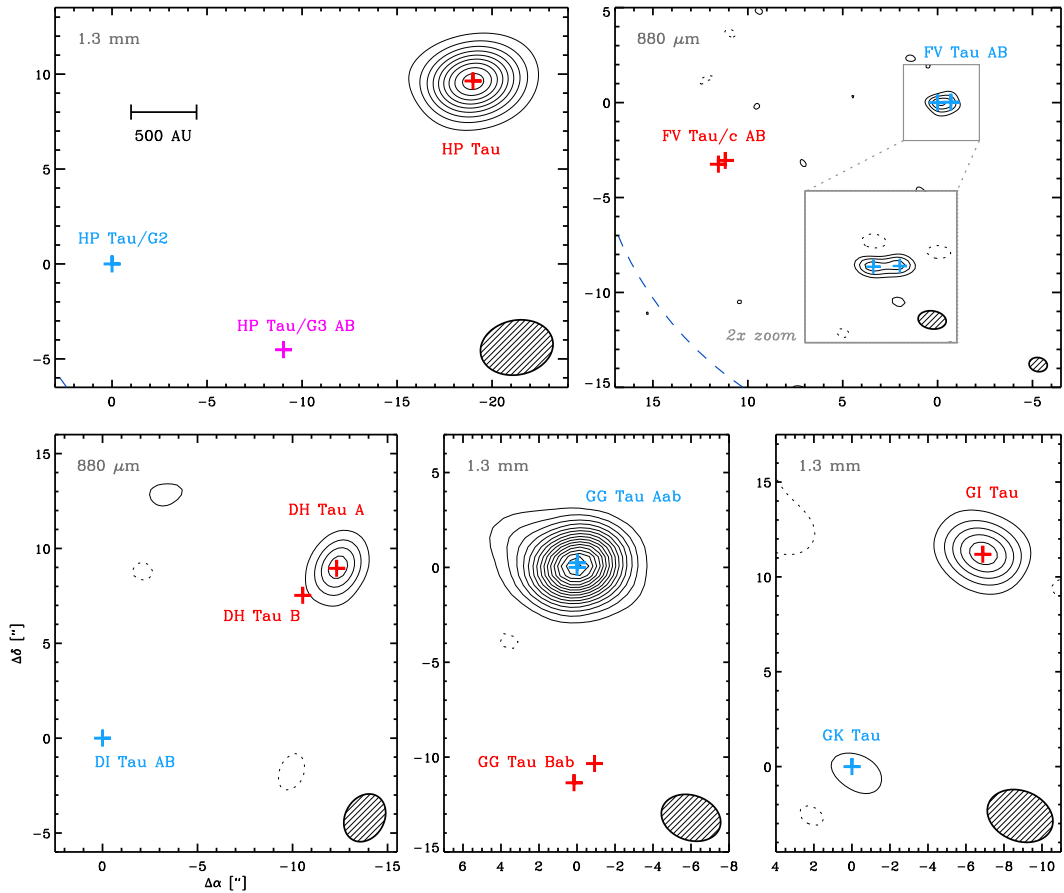}
\figcaption{Mosaic of SMA continuum images for wide-separation multiple
systems.  Contours are drawn at 3\,$\sigma$ intervals (see Table
\ref{table:maps}); hatched ellipses mark the synthesized beam dimensions.
Stellar positions are marked with blue (primary), red (secondary), or purple
(intermediate mass in higher-order systems) crosses.  A blue dashed curve marks
the SMA primary beam in the FV Tau map.  The FV Tau inset shows the data 
synthesized at higher resolution, with a beam size of $0\farcs7 \times 
0\farcs5$.  The panels are drawn to scale; a 500\,AU scale bar is marked in the 
HP Tau system panel ({\it top left}).  \label{fig:images1}}
\end{figure}

In most cases, the SMA dual-sideband receivers were tuned to a local oscillator 
(LO) frequency of 340.755\,GHz (880\,$\mu$m) or 225.497\,GHz (1.3\,mm).  Some 
tracks used shifted LO settings to accomodate other projects that shared one 
night of observing.  The data obtained in 2010-2011 employs two IF bands (per 
sideband) spanning $\pm$4-6\,GHz and and $\pm$6-8\,GHz from the LO frequency 
(only the lower IF band was available for the 2 observations in 2005 and 
2009).  Each IF band contains 24 partially overlapping 108\,MHz-wide spectral 
chunks (per sideband).  Aside from one chunk reserved for the local CO 
transition, each of these was coarsely divided into 32 channels to observe the 
continuum.  A finer sampling of 256 channels per chunk was used to probe the CO 
emission, corresponding to a velocity resolution of 0.40 and 0.55\,km s$^{-1}$ 
near the $J$=3$-$2 and $J$=2$-$1 transitions, respectively.  The observations 
cycled between various target systems and the nearby quasars 3C 111 and 
J0510+180 on timescales of $\sim$20 minutes for the compact array and 
10\,minutes for the longer baseline configurations.  Bright quasars (3C 279 or 
3C 454.3), Uranus, and satellites (Titan, Callisto) were observed as bandpass 
and absolute amplitude calibrators when the targets were at low elevations.  
Observing conditions were often excellent, with precipitable water vapor levels 
ranging from 1.0-1.6\,mm and $\sim$2-3\,mm for the 880\,$\mu$m and 1.3\,mm 
observations, respectively.  

\begin{figure}
\epsscale{0.95}
\plotone{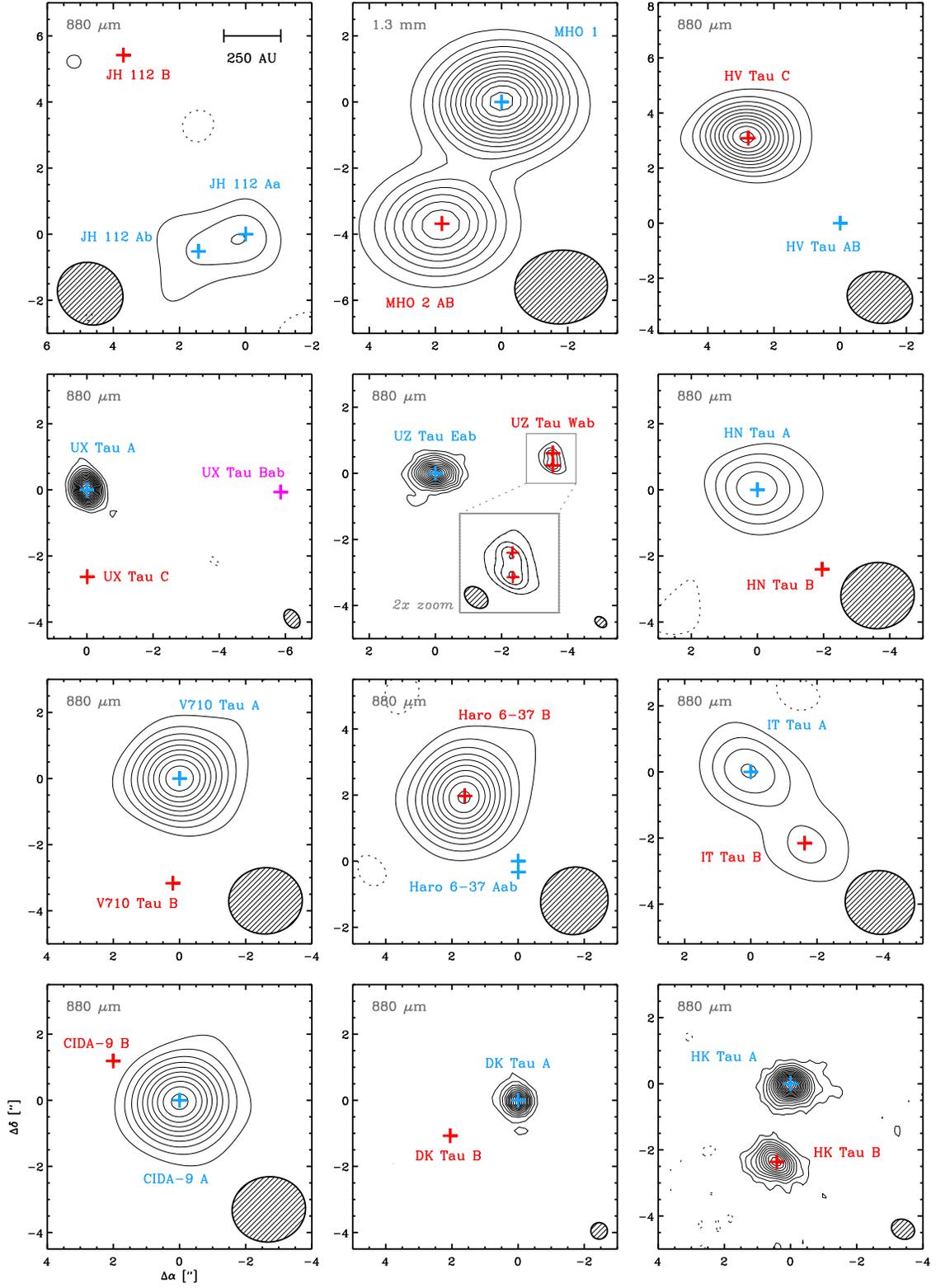}
\figcaption{Same as Fig.~\ref{fig:images1}, for medium separations.  
\label{fig:images2}}
\end{figure}

\begin{figure}[ht]
\epsscale{0.9}
\plotone{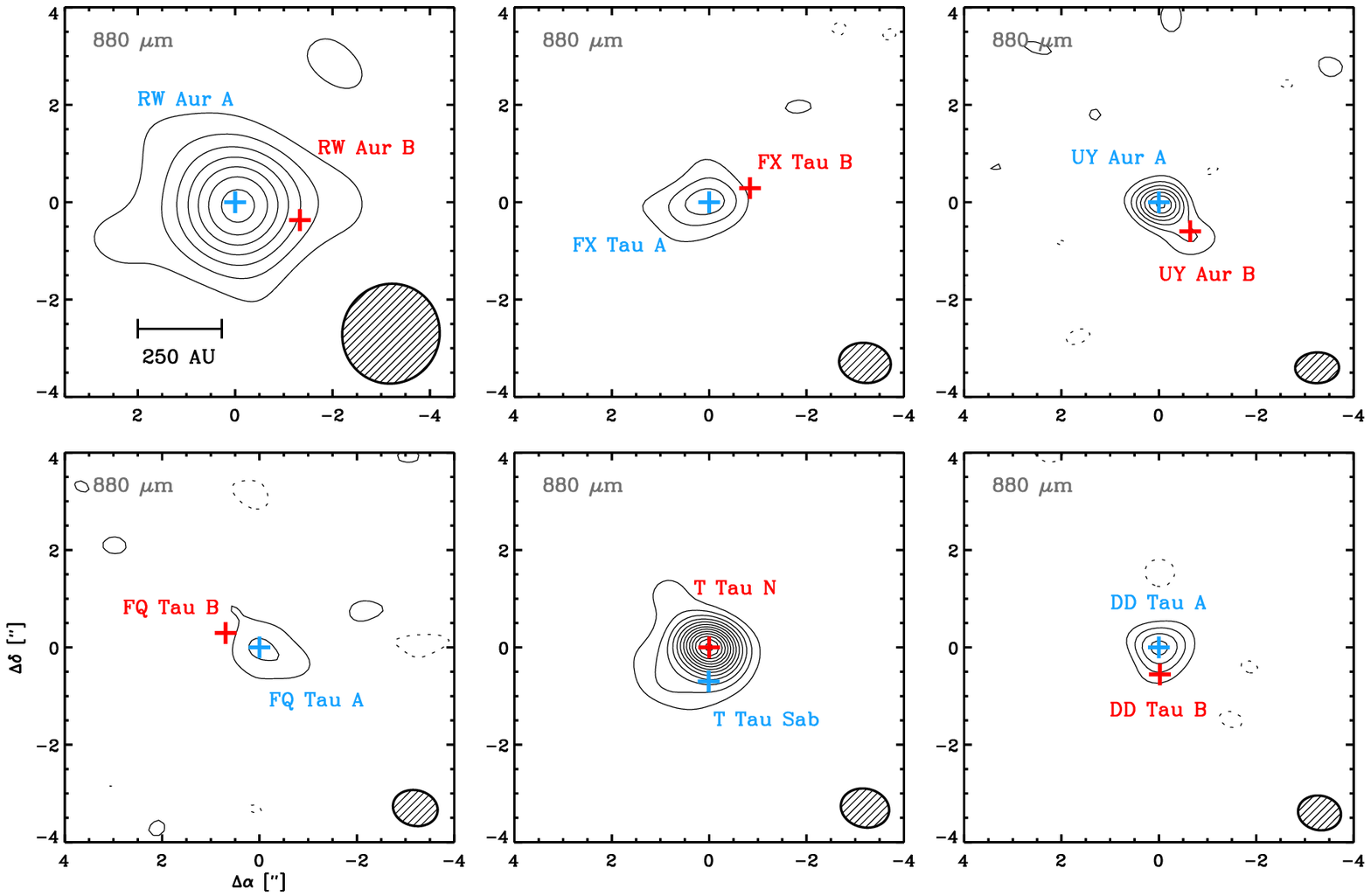}
\figcaption{Same as Figs.~\ref{fig:images1} and \ref{fig:images2}, for small 
separations.  \label{fig:images3}}
\end{figure}

The data were reduced with the IDL-based {\tt MIR} software package.  The 
spectral response was calibrated using observations of bright quasars as 
references, and the central 82\,MHz from the individual spectral chunks in each 
sideband and IF band were averaged into an effective continuum channel 
(excluding the chunk containing a CO transition).  The antenna-based complex 
gain response of the system was determined using the phase calibrator 
nearest to the target.  The absolute amplitude scale was set based on 
observations of Uranus or planetary satellites, and is expected to be accurate 
at the level of $\sim$10\%.  For each target, the continuum channels for both 
IF bands and sidebands from each set of observations were combined into a 
composite set of calibrated visibilities.  The {\tt MIRIAD} software package 
was used to Fourier invert those visibilities, perform a deconvolution using 
the {\tt CLEAN} algorithm, and restore the {\tt CLEAN}ed maps with a 
synthesized beam.  The synthesized beam dimensions and RMS noise levels for the 
naturally weighted datasets are provided in Table \ref{table:maps}.  The SMA 
continuum maps are shown in Figures \ref{fig:images1}-\ref{fig:images3}.  In 
most cases, the observations of a given multiple system were not sufficient to 
clearly detect CO emission from any circumstellar gas: the few exceptions will 
be discussed elsewhere.

\section{Disk Properties from Simple Emission Models}

The SMA survey observations described above comprise the largest resolved 
millimeter-wave census of circumstellar material in young multiple star systems 
to date.  In this section, we aim to measure two fundamental properties from 
these data -- luminosities (which are related to dust masses) and sizes -- for 
the disks around the individual stellar components in each multiple system.  
These basic disk parameters are estimated by fitting a simple model of the 
continuum emission morphology directly to the observed visibilities.  The 27 
multiple systems in our sample contain 77 individual stars.  The available data 
have sufficient angular resolution to associate any dust emission with 50 of 
those stars.  The individual components in the close pairs MHO 2 AB, T Tau Sab, 
FS Tau AB, DI Tau AB, UX Tau Bab, XZ Tau AB, GG Tau Aab, UZ Tau Eab, V807 Tau 
AB, GH Tau AB, HP Tau/G3 AB, HV Tau AB, and Haro 6-37 Aab are not resolved.  We 
treat the millimeter signal from each of these 13 pairs as if it arises from a 
``composite" disk (see \S 5.1).  

Including two of those composites, there are 14 individual disks in this sample 
that are sufficiently well-resolved to provide robust estimates of their basic 
parameters.  In these cases, we define a simple, azimuthally symmetric and 
geometrically flat emission model with a power-law radial surface brightness 
distribution, $I_{\nu} \propto R^{-x}$, that extends to an outer edge, $R_d$.  
The emission profile is normalized such that the total flux density $F_d = \int 
I_{\nu} d\Omega$.  This parametric emission morphology is designed to mimic 
what would be expected from a disk structure model with power-law surface 
density ($\Sigma_d \propto R^{-p}$) and temperature ($T_d \propto R^{-q}$) 
profiles.  Pressing that resemblance, the radial index $x$ is analogous to the 
sum $p+q$ and the normalization $F_d$ is a rough proxy for the product 
$\kappa_d \langle T_d \rangle M_d$, where $\kappa_d$ is the dust opacity, 
$\langle T_d \rangle$ is a characteristic temperature, and $M_d$ is the dust 
mass, modulo small correction factors for any high optical depths in the disk 
center \citep{beckwith90,aw05}.  Our data do not generally have enough 
sensitivity on long baselines to provide useful quantitative constraints on 
both the emission gradient and size, which are effectively degenerate at this 
modest resolution \citep[see][]{mundy96,aw07b}.  Since the key parameters of 
interest from the perspective of tidal interaction models are \{$F_d$, $R_d$\}, 
we elect to fix the gradient to a fiducial value, $x = 1.5$, motivated by the 
standard assumptions for irradiated accretion disks \citep[$p = 1$, $q = 0.5$; 
see][]{hartmann98}.  For reference, adjustments to the radial index of 
$\pm$30\%\ ($\Delta x \approx \pm 0.5$) induce systematic changes in the size 
estimates of $\sim$20-40\%: steeper (shallower) gradients produce larger 
(smaller) sizes.

\begin{figure}[t]
\epsscale{0.85}
\plotone{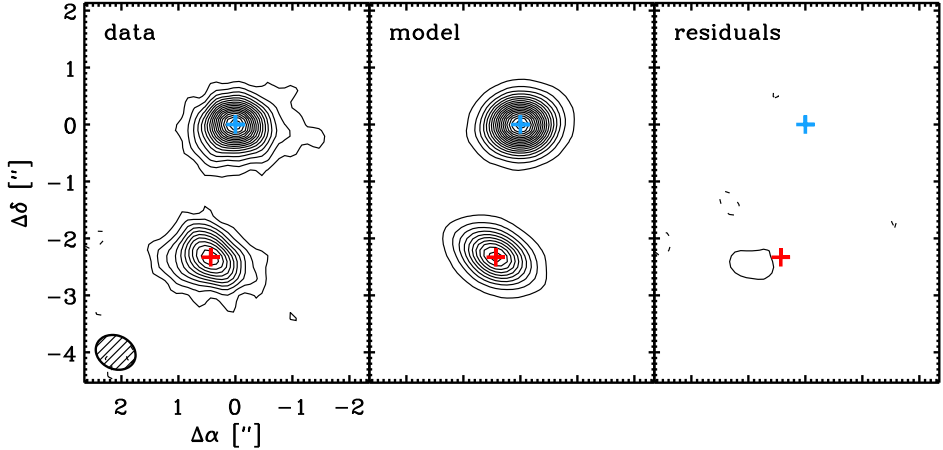}
\figcaption{Model fit demonstration for the HK Tau binary.  From left to right, 
the panels show the SMA 880\,$\mu$m image, the best-fit model image, and the 
imaged residuals.  Contours are drawn at 3\,mJy beam$^{-1}$ (3\,$\sigma$) 
intervals.  The synthesized beam is shown in the lower left of the data panel.  
The HK Tau A and B stellar positions are marked with blue and red crosses, 
respectively.  \label{fig:fit_demo}}
\end{figure}

In addition to the two free parameters in the surface brightness model, 
\{$F_d$, $R_d$\}, there are formally five other parameters related to the 
projection of the model into the sky plane: the disk center relative to the 
observed phase center \{$\Delta \alpha$, $\Delta \delta$\}, the disk viewing 
geometry described by its apparent inclination and orientation \{$i_d$, 
PA$_d$\}, and the distance to the observer \{$d$\}.  The latter is fixed to $d 
= 145$\,pc, with a systematic uncertainty estimated to be roughly $\pm$10\%\ 
\citep{loinard07,torres07,torres09}.  We fix the centroid positions before 
estimating other parameters, typically based on an elliptical Gaussian fit to 
the visibilities for individual components that exhibit continuum emission.  In 
general, that technique recovers the expected stellar positions well within the 
position accuracy of the SMA data ($\sim$0\farcs1 in an absolute sense, and 
considerably better in a relative sense for the few cases with multiple disk 
detections).  For stellar components that do not exhibit any millimeter 
emission or may be partially blended with other components, we rely on the 
positions (or projected angular separations and orientations) provided from 
optical/infrared measurements in the literature to assign their \{$\Delta 
\alpha$, $\Delta \delta$\} values.  In practice, we estimate the best-fit 
values of 4 free parameters \{$F_d$, $R_d$, $i_d$, PA$_d$\} and their 
uncertainties for each resolved disk by comparing model predictions directly 
with the SMA visibilities using the non-linear $\chi^2$ minimization routine 
{\tt MPFIT} \citep{markwardt09}.  In each case, several randomized initial 
parameter sets were employed to avoid trapping in local minima.  The results 
are compiled in Table \ref{table:modelfits}.  Figure \ref{fig:fit_demo} shows 
an example model fit for the well-separated and resolved disks of the HK Tau 
binary.  

For 4 of the 14 disks with modeling results in Table \ref{table:modelfits}, the 
emission is not sufficiently resolved to place meaningful constraints on the 
disk viewing geometry.  In those cases (FX Tau A, UZ Tau Wb, HN Tau A, and 
Haro 6-37 B), we assumed a fiducial \{$i_d = 45\degr$, PA$_d = 90\degr$\} in 
order to calculate a reasonable estimate of \{$F_d$, $R_d$\} (alternative 
viewing geometry selections produce the same flux densities and sizes within 
the quoted uncertainties).  The viewing geometries for the HK Tau B, HV Tau C, 
and GG Tau Aab disks were fixed based on observations of their scattered light 
morphologies or molecular kinematics \citep{duchene10,mccabe11,guilloteau99}.  
The UX Tau A and GG Tau Aab disks were modeled as rings, with empty central 
regions out to radii of 25 and 185\,AU based on the more sophisticated analyses 
of \citet{andrews11} and \citet{guilloteau99}, respectively.  The UZ Tau Eab 
circumbinary disk was also modeled as a ring, with no emission inside a radius 
of 15\,AU: a detailed analysis of this disk will be provided elsewhere (Harris 
et al., in preparation).  Of the remaining 49 individual stars or close-pair 
composites, 27 have firmly detected -- but {\it unresolved} -- millimeter 
emission and 22 others do not.  For simplicity, point source models were used 
to measure $F_d$ for the population of detected, but unresolved, disks.  After 
subtracting models of any emission from nearby stars, upper limits 
(3\,$\sigma$) on $F_d$ were estimated for the undetected components by 
computing the RMS noise level in a $4\arcsec\times4\arcsec$ box centered on the 
stellar position in a synthesized residual map.  The point source flux 
densities and upper limits for these 49 individual stellar components (or 
close-pair composites) are compiled in Table \ref{table:modelfits_unresolved}.  

The emission from the individual disks in the RW Aur and DD Tau binaries is 
blended.  In these cases, we adopted an iterative modeling strategy.  First, a 
single disk model was fitted to the dominant emission component (the primaries) 
and subtracted from the data.  An initial estimate of the disk model for the 
blended component (the secondaries) was made from the residuals.  Based on 
those results, a composite model with both disk components was then used to fit 
the data and derive proper parameter estimates and uncertainties.  This method 
naturally accounts for the blending with inflated formal parameter 
uncertainties for each model component.  None of the disks in these pairs is 
spatially resolved, so the underlying model for each component disk is a point 
source.

\section{Results}

We aim to take advantage of this resolved millimeter-wave census to address 
some key aspects of disk evolution in the presence of stellar companions.  To 
do that, we link our SMA survey results into a comprehensive compilation of 
stellar separations ($\rho$), component masses ($M_p$ and $M_s$) and mass 
ratios ($q$), and millimeter luminosities (defined as the summed emission in a 
pair: $F_{\rm pair} = F_{d,p}+F_{d,s}$) for {\it all} of the known potentially 
interacting stellar pairs in Taurus (with spectral types F0 to M4).  Those data 
are compiled in Table \ref{table:pairs}.  A complementary list of {\it single} 
stars in Taurus with available millimeter-wave observations in the literature 
is provided in Table \ref{table:singles}.  We should note that there are 36 
other single stars in this spectral type range in the compilation of 
\citet{luhman10} that, to our knowledge, have not yet been observed at 
millimeter wavelengths.

\subsection{Millimeter Detection Statistics}

If interactions with companions efficiently remove material from circumstellar 
disks in young multiple systems, there should be a clear signature in the 
relative detection fractions of millimeter-wave emission (a rough proxy for 
dust mass) between isolated (single) stars and the individual stellar 
components of multiple systems: the fraction of stars that exhibit detectable 
millimeter emission, $f_{\rm mm} = N_{\rm det} / N_{\rm tot}$, should be 
substantially higher for singles compared to multiples.  This feature has been 
noted anecdotally in the past 
\citep[e.g.,][]{jensen94,jensen96,osterloh95,aw05}, but the inability to assign 
millimeter emission to individual components in multiple systems has limited 
any firm quantitative assessment of the detection statistics.  

There are millimeter-wave continuum measurements available for 52 single stars 
in Taurus, 48 binaries (96 stars), 13 triples (39 stars), 7 quadruples (28 
stars), 2 quintuples (10 stars), and 1 sextuple (6 stars; the LkH$\alpha$ 332 
system).  Using the component-resolved $F_d$ measurements in Tables 
\ref{table:modelfits} and \ref{table:modelfits_unresolved} along with the 
additional literature photometry compiled in Tables \ref{table:pairs} and 
\ref{table:singles}, we computed the $f_{\rm mm}$ values for single stars and 
multiples listed in Table \ref{table:fmm}.  The ranges of $N_{\rm det}$ and 
$f_{\rm mm}$ for multiples correspond to the potential distribution of the 
observed millimeter emission among any unresolved components.  We find that 32 
of 52 single stars show millimeter emission, $f_{\rm mm} = 62\pm11$\%, while 
only 50-67 of 179 individual stars in multiple systems have millimeter-wave 
detections, $f_{\rm mm} = 28$-37$\pm5$\%.  A two-tailed Fisher Exact test 
confirms that these are indeed statistically different detection fractions, 
with a $p$-value $<0.002$.  Remarkably, the millimeter detection fraction for 
individual stars in multiple systems does not depend on the number of 
companions: binaries, triples, and higher-order groups all have $f_{\rm mm} 
\approx 1/3$, roughly half the detection rate for singles.  That uniformity is 
a testament to the hierarchical nature of multiples, where higher-order 
($N_{\ast} \ge 3$) systems tend to be be constructed of sets of binary pairs.

Given that the detection fraction for stars with companions is roughly half 
that for stars without them, it may seem natural to assume that only one 
stellar component in a multiple retains disk material.  While not uncommon, 
this is not necessarily the typical scenario.  Of the 48 binaries in Taurus, 20 
exhibit millimeter emission.  Of those 20 pairs, our millimeter observations 
have resolved the individual components of 12 (from 24 stars).  In 6 of those 
pairs, the emission is concentrated solely around the primary.  In the other 6, 
both components show some emission -- however, the primary is {\it always} 
brighter.  So, for the sub-population of component-resolved binaries with 
millimeter emission, the detection fraction for individual stars is actually 
fairly large (18/24).  In the higher order multiples, the situation is slightly 
more complicated by their hierarchical structure.  In some cases, we find dust 
emission coincident with all components, albeit usually with some pair 
presumably surrounded by a circum{\it binary} disk (e.g., UZ Tau, MHO 1/2).  In 
others, the dust emission is only present around a more isolated, distant 
companion (e.g., HV Tau, Haro 6-37).  

As might be expected, the likelihood that any individual component of a 
multiple system harbors a circumstellar disk that is massive enough to generate 
detectable millimeter emission depends critically on the individual details of 
the system.  The following sections explore the potentially observable 
signatures expected from star-disk interactions, with a more explicit focus on 
some key connections between the stellar properties tied to orbital dynamics 
and the basic disk characteristics that can be inferred from the millimeter 
data.

\subsection{Pair Demographics and Disk -- Star Connections}

Theoretical models of star-disk interactions in binary pairs suggest that the 
separation between the stellar components is the key property that controls 
the tidal truncation of individual circumstellar disks 
\citep{al94,pichardo05}.  These interactions effectively remove mass from the 
outer regions of these disks (either by accretion or ejection into the local 
interstellar medium), such that stellar pairs with smaller separations should 
harbor smaller -- and therefore less massive -- disks.  Since the cool dust in 
the outer disk that is being stripped by this process emits continuum radiation 
at millimeter wavelengths, these tidal interactions should naturally produce an 
observable trend where the millimeter luminosity is positively correlated with 
the separation of the stellar pair.  Indeed, the pioneering work on this 
subject by \citet{jensen94,jensen96} clearly identified that the millimeter 
luminosities for pairs with projected separations $a_p \le 50$-100\,AU were 
statistically lower than their more widely-separated counterparts or single 
stars.  These different populations were confirmed with larger and deeper 
millimeter-wave surveys \citep{osterloh95,aw05}, but the detailed distribution 
of $F_{\rm pair}$ with respect to $a_p$ has remained unclear for two reasons: 
the low resolution of single-dish millimeter-wave photometry often included 
several pairs together, and the multiplicity census of the nearest star-forming 
regions was incomplete.  Our survey mitigates these issues, providing an 
opportunity to look at the details of the millimeter luminosity--separation 
distribution.

\begin{figure}[t]
\epsscale{1.1}
\plottwo{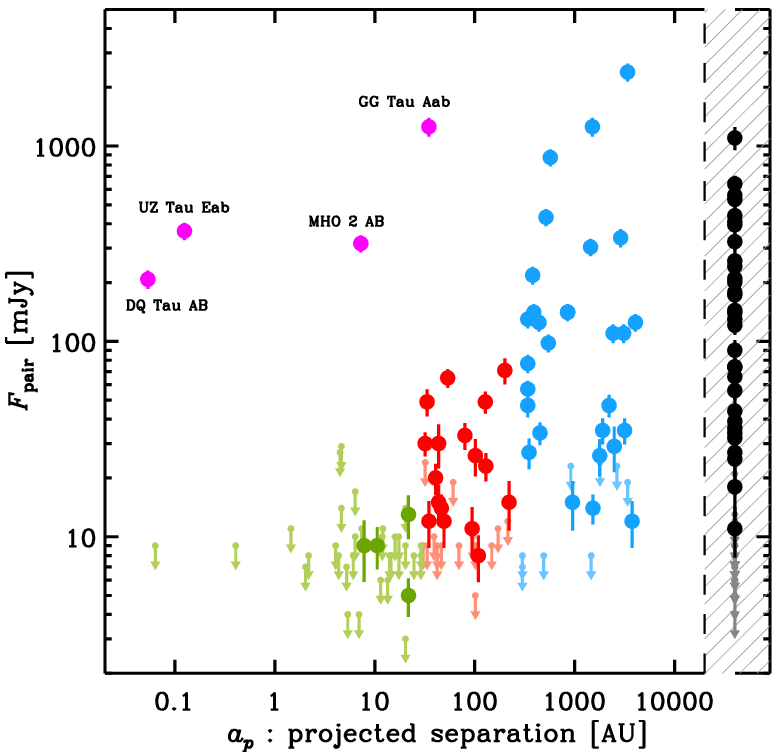}{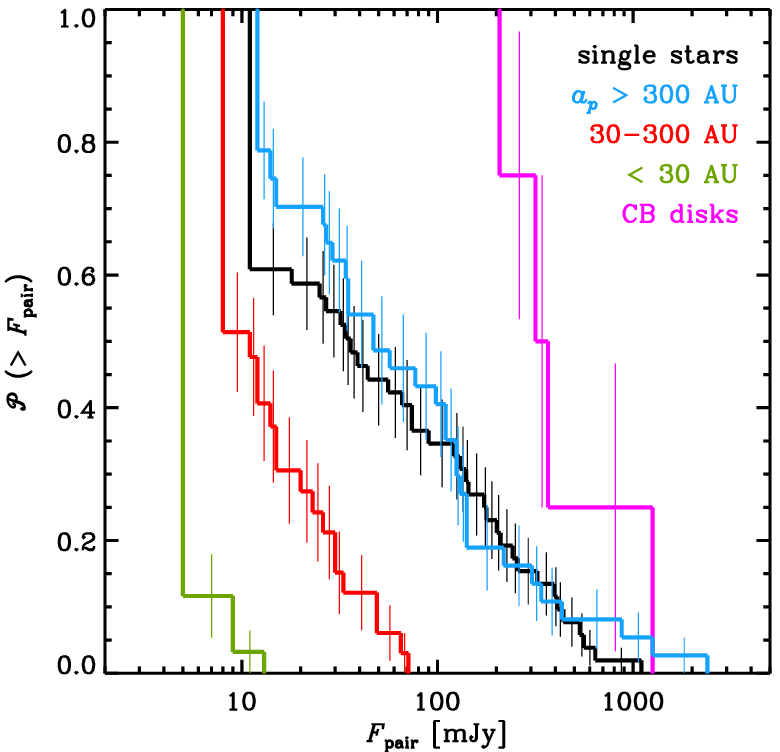}
\figcaption{($a$) A comparison of millimeter flux densities from potentially 
interacting pairs as a function of the projected pair separation.  Single stars 
are shown to the right of the plot as black points for reference.  The pair 
population can be distinguished into 4 clear sub-categories: wide ($a_p > 
300$\,AU), medium ($a_p = 30$-300\,AU), and small ($a_p < 30$\,AU) pairs, and 
circumbinary disks ({\it purple}).  ($b$) The cumulative distributions of 
millimeter flux densities for each of these sub-categories, constructed with 
the Kaplan-Meier product-limit estimator to include the available upper 
limits.  Millimeter luminosity is strongly dependent on projected separation. 
\label{fig:flux_sep}}
\end{figure}

Figure \ref{fig:flux_sep}$a$ shows $F_{\rm pair}$ as a function of $a_p$ for 
the 111 stellar pairs in Taurus (see Table \ref{table:pairs}), along with the 
52 single stars that have millimeter-wave measurements (see Table 
\ref{table:singles}; black points, gray upper limits).  This diagram is a 
striking confirmation of the original conclusions of Jensen et al., plainly 
demonstrating that millimeter continuum luminosities scale with the separation 
between stellar pairs.  However, as the pair separation decreases, the {\it 
maximum} millimeter luminosities appear to decline in discrete jumps (rather 
than continuously) at two relatively well-defined locations: $a_p \approx 300$ 
and 30\,AU ($\rho \approx 2$ and 0.2\arcsec, respectively).  These features 
facilitate a natural breakdown of the Taurus pairs into distinct 
sub-populations.  The distribution of $F_{\rm pair}$ for wide pairs, with 
projected separations greater than 300\,AU ({\it blue}), is similar to the 
distribution for single stars.  At medium separations -- $a_p = 30$-300\,AU 
({\it red}) -- we find a notable absence of bright pairs and a decreased 
detection rate.  At yet smaller separations, $a_p < 30$\,AU ({\it green}), only 
a few pairs exhibit very weak millimeter-wave emission.  A small group of 
dramatic outliers sparsely populate the otherwise empty region of bright pairs 
with small separations ({\it purple}): these pairs are known or suspected to 
harbor massive circum{\it binary} rings \citep[e.g., see][regarding GG Tau 
Aab]{pietu11}.  

The cumulative distributions of $F_{\rm pair}$ for each separation-based 
sub-population can better quantify this apparent trend.  The ${\mathcal 
P}_a(>$$F_{\rm pair})$ distributions shown together in Figure 
\ref{fig:flux_sep}$b$ were constructed using the Kaplan-Meier product-limit 
estimator to properly account for the substantial number of pairs in each 
sub-population that exhibit no millimeter emission \citep[see][]{feigelson85}.  
The distributions are compared directly with the standard two-sample tests used 
in survival analysis in Table \ref{table:surv_tests}.  These tests confirm the 
qualitative examination of Figure \ref{fig:flux_sep}$a$: (1) wide pairs and 
single stars have statistically indistinguishable millimeter luminosity 
distributions; (2) medium pairs have significantly lower luminosities; and (3) 
small pairs have yet less millimeter emission.  The ${\mathcal P}_a(>$$F_{\rm 
pair})$ in Figure \ref{fig:flux_sep}$b$ have similar functional forms, albeit 
shifted in luminosity.  A simple scaling indicates that $F_{\rm pair}$ 
decreases by a factor of $\sim$5 from wide to medium separations, and then 
another factor of 5 from medium to small separations.  These trends are not an 
artifact of including non-detections in the analysis: similar conclusions are 
drawn by comparing only the detected pairs in the same sub-populations using 
the two-sided Kolmogorov-Smirnov (K-S) test (see Table \ref{table:surv_tests}). 

\begin{figure}[t]
\epsscale{1.1}
\plottwo{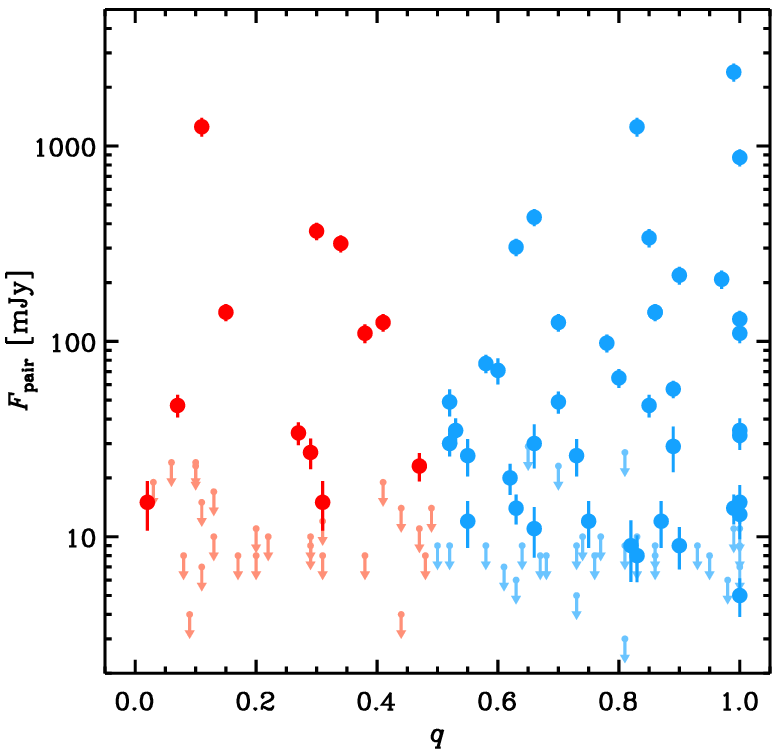}{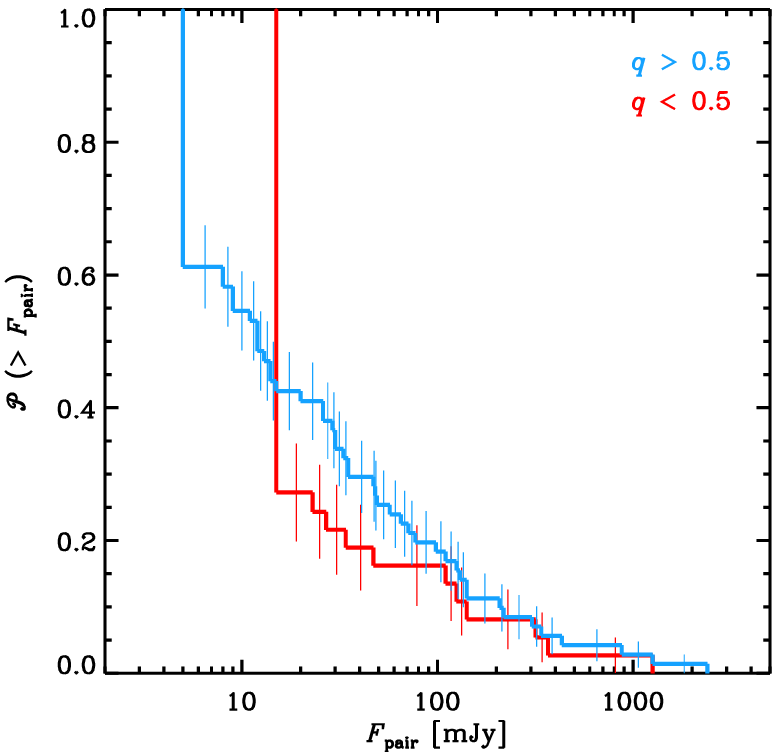}
\figcaption{($a$) A comparison of millimeter flux densities from potentially
interacting pairs as a function of stellar mass ratio.  For reference, the pair
population is distinguished into high ($q > 0.5$; {\it blue}) and low ($q < 
0.5$; {\it red}) mass ratio groups.  ($b$) The cumulative distributions of
millimeter flux densities for high and low mass ratio pairs, constructed with
the Kaplan-Meier estimator to include the available upper limits.  Millimeter
luminosity depends only weakly on the stellar mass ratio.  \label{fig:flux_q}}
\end{figure}

Models of star-disk interactions also postulate an association between the 
amount of disk truncation and the component masses of the stellar pair.  
Massive companions impart larger dynamical perturbations to individual disks, 
producing more tidal stripping, and leading to lower disk masses and therefore 
less millimeter-wave emission.  In that case, $F_{\rm pair}$ should be 
anti-correlated with $q$: higher mass ratio pairs should have fainter disk 
emission.  No such trend is obvious in Figure \ref{fig:flux_q}$a$.  If we 
separate the full population into high and low mass ratio pairs at some 
critical $q_c$, we find the largest difference between those sub-categories for 
$q_c = 0.5$.  Figure \ref{fig:flux_q}$b$ compares the cumulative distributions 
of the millimeter flux densities for high ($q > 0.5$; {\it blue}) and low ($q < 
0.5$; {\it red}) mass ratio pairs, again with $\mathcal{P}_q(>F_{\rm pair})$ 
constructed using the Kaplan-Meier estimator to include the pairs that do not 
have detectable millimeter emission.  The same two-sample tests employed above 
indicate a weak relationship between $F_{\rm pair}$ and $q$ (see Table 
\ref{table:surv_tests}), such that pairs with low stellar mass ratios have 
slightly less millimeter emission -- the opposite of expectations from tidal 
interaction models.  However, the evidence for any increase in the millimeter 
emission with higher stellar mass ratios is contained entirely in the relative 
detection ratios: a larger fraction of pairs with low $q$ are not detected at 
millimeter wavelengths (see Figure \ref{fig:flux_q}$a$).  If only the detected 
pairs are compared, the millimeter luminosity is found to be independent of $q$ 
(see the K-S test results in Table \ref{table:surv_tests}).  Moreover, this 
trend is present (and in fact enhanced) only for stellar pairs with wide 
separations: no clear relationship between $F_{\rm pair}$ and $q$ exists for 
pairs with $a_p < 300$\,AU (see Table \ref{table:surv_tests}).

\begin{figure}[t]
\epsscale{1.1}
\plottwo{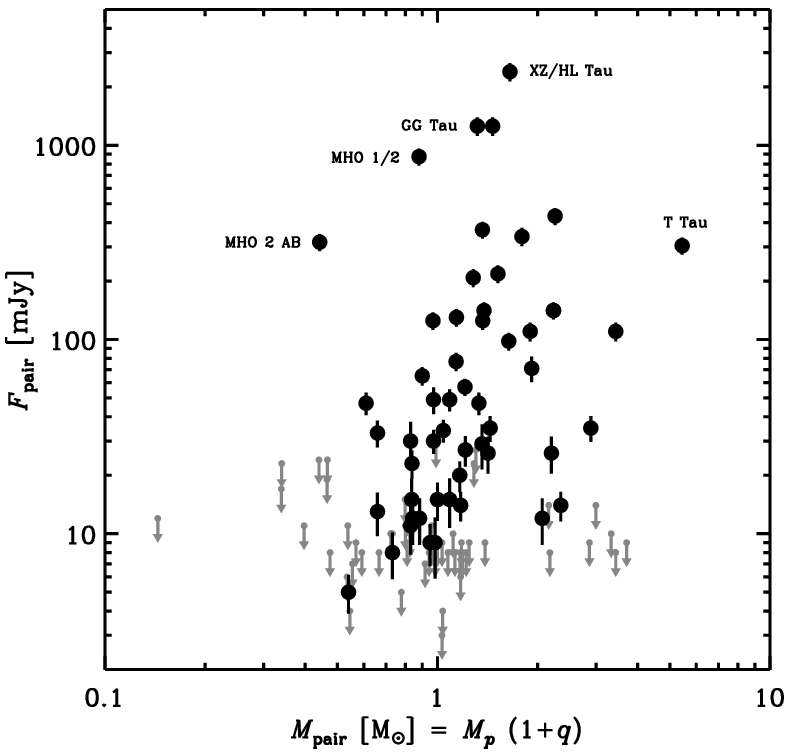}{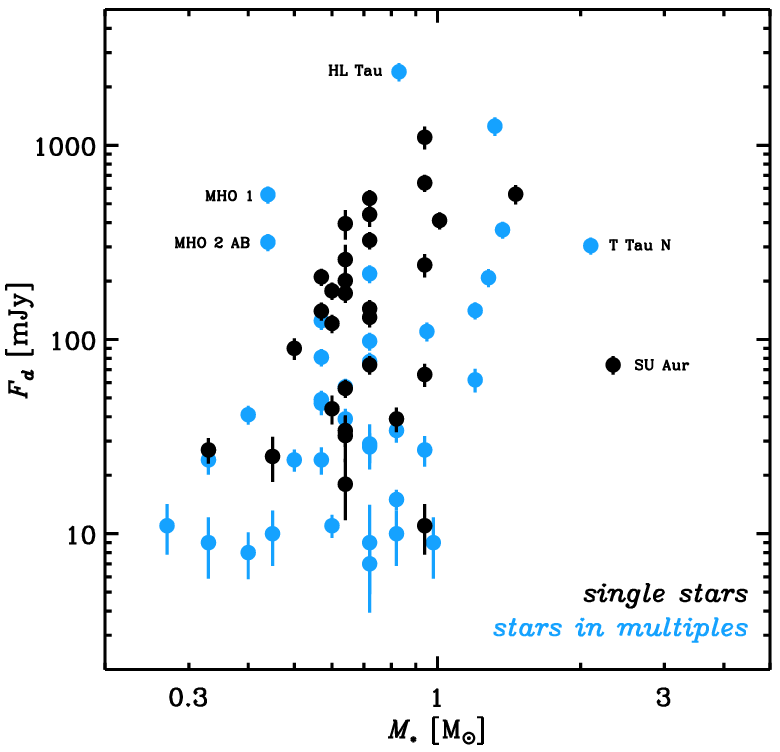}
\figcaption{($a$) Millimeter flux density as a function of the total mass of a 
stellar pair, where $M_{\rm pair} = M_p+M_s = M_p(1+q)$.  ($b$) Millimeter flux 
densities as a function of stellar mass for the individual, detected stellar 
components in Taurus, for both single stars ({\it black}) and those in multiple 
systems ({\it blue}).  Note that upper limits for non-detections are not shown 
in the right-hand panel.  \label{fig:flux_m}}
\end{figure}

The absence of a firm connection between the millimeter luminosity and stellar 
mass ratio for smaller-separation pairs is consistent with the weak 
$q$-dependence predicted by tidal interaction models.  Nevertheless, the 
correlation between $F_{\rm pair}$ and $q$ for widely-separated pairs is 
compelling.  Since star-disk interactions in these cases should be inherently 
less destructive, this trend might indicate a relationship between the 
millimeter luminosity and the absolute (and not relative) stellar masses in the 
pair.  Indeed, Figure \ref{fig:flux_m}$a$ demonstrates a marginal association 
between the millimeter luminosity and total stellar mass in Taurus pairs.  
Taking only the detections, we find a $\sim$3\,$\sigma$ correlation between 
$F_{\rm pair}$ and $M_{\rm pair}$ (this improves slightly to 3.7\,$\sigma$ if 
the labeled outliers are excluded).  Figure \ref{fig:flux_m}$b$ demonstrates 
that this relationship is not restricted to stellar pairs, but apparently also 
applies to {\it individual} stars, both isolated (single) cases and members of 
multiple systems.  In the latter case, we have combined the component-resolved 
$F_d$ estimates from Table \ref{table:modelfits} and the single star 
measurements from the literature (Table \ref{table:singles}): the result is 
again a $\sim$3\,$\sigma$ positive correlation (4.5\,$\sigma$ if SU Aur is 
excluded).  It is worth noting that the $M_{\ast}$ distributions for isolated 
stars and individual stars in multiple systems are statistically 
indistinguishable.  However, it is premature to draw any firm conclusions from 
these weak trends for two reasons: (1) the dispersion is large relative to 
the range of the trend, and (2) the stellar masses used here have large 
systematic uncertainties.  

\begin{figure}[t]
\epsscale{1.1}
\plottwo{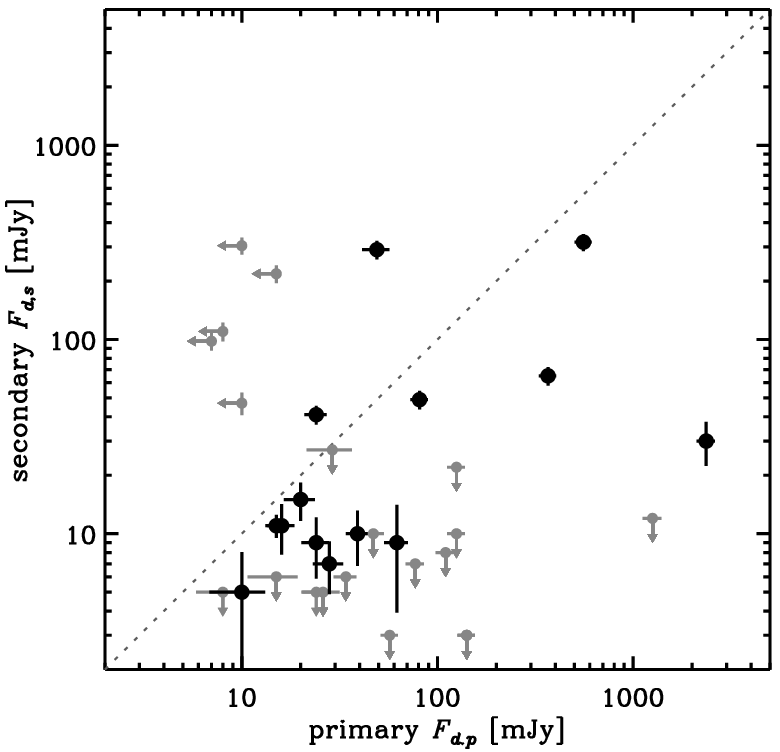}{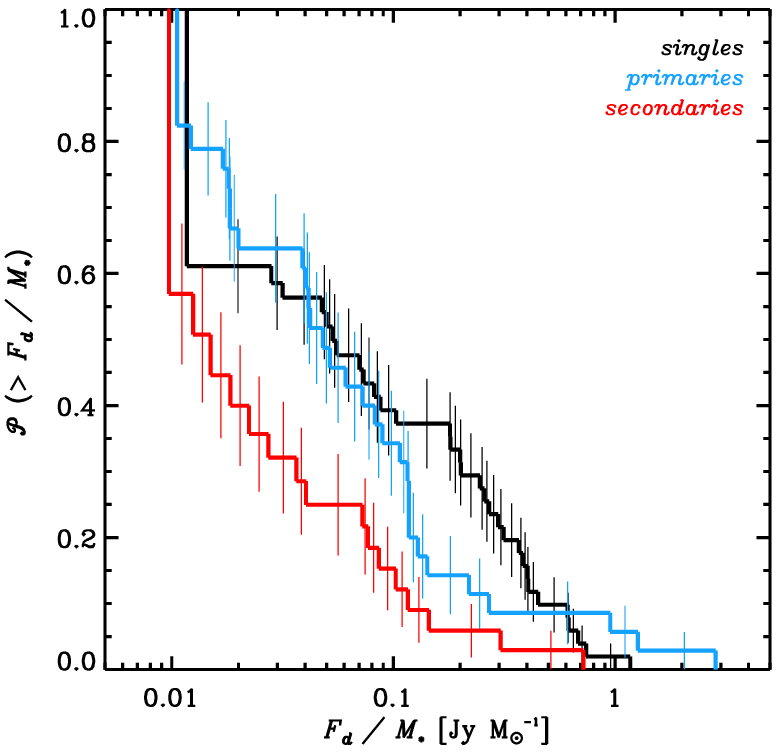}
\figcaption{($a$) The 880\,$\mu$m flux densities for the primary and secondary 
components of each stellar pair.  ($b$) The cumulative distributions of the 
ratio of millimeter-wave disk flux densities to their stellar host masses (a 
proxy for the disk:star mass ratio) for singles ({\it black}), primaries ({\it 
blue}), and secondaries ({\it red}), each constructed from the Kaplan-Meier 
product-limit estimator to include upper limits on $F_d$.  Two-sample tests 
(see Table \ref{table:surv_tests}) demonstrate that these ratios are 
systematically lower for secondaries, implying that their disks are inherently 
less luminous regardless of the host masses.\label{fig:fratios}}
\end{figure}

The demographic properties of stellar pairs discussed above are certainly 
informative, but they also naturally hide some characteristics of individual 
components that are available from this resolved survey.  The two panels in 
Figure \ref{fig:fratios} are intended to compare the millimeter-wave 
luminosities from individual disk components {\it within} each stellar pair, 
as a function of their stellar host masses.  Figure \ref{fig:fratios}$a$ 
directly compares the resolved 880\,$\mu$m flux densities for each component in 
each pair.  As mentioned in \S 5.1, the emission from the primary is usually 
more luminous than from the secondary.  The exceptions above the dashed line 
are comprised of the widely-separated tertiary companions of close pairs and 
the UZ Tau Wab and FS Tau/Haro 6-5B pairs.  The relative dominance of the 
primary disk emission is unaffected by the projected separation to a companion, 
and is typically more than would be expected if the amount of millimeter-wave 
disk emission scales linearly with the stellar host mass.  This latter feature 
is shown more directly in Figure \ref{fig:fratios}$b$, which compares the 
cumulative distributions of the ratio of the millimeter flux density to the 
stellar mass (akin to a disk:star mass ratio) for singles, primaries, and 
secondaries.  In some studies of disks around low-mass stars and brown dwarfs, 
this $F_d/M_{\ast}$ ratio is found to be roughly constant; the weaker emission 
(or lower detection rate) is a manifestation of inherently lower host masses 
\citep[e.g.,][]{scholz06,schaefer09}.  That implicit sensitivity threshold is 
not the case for the secondaries in our sample: the $F_d/M_{\ast}$ ratio is 
systematically lower for secondaries compared to primaries and singles, with a 
probability that it is drawn from the same parent distributions as the 
primaries or singles of $<$0.008 (two-sample test results are compiled in Table 
\ref{table:surv_tests}).  The results suggest that the millimeter-wave disk 
emission from the secondaries is inherently less luminous than around the 
primaries (or isolated stars), regardless of the stellar host mass.

\subsection{Disk Sizes and Tidal Truncation}

A more direct test of dynamical predictions for star-disk interactions lies 
with our measurements of individual disk sizes.  Theoretical models provide a 
way to estimate the truncated equilibrium tidal radii ($R_t$) for the disks 
around each component of a stellar pair given a few key orbital parameters, 
\{$a$, $e$, $q$\}, and some characterization of the viscous properties of the 
disk material \citep{al94}.  Although they remain uncertain, there are 
reasonable ways to estimate stellar mass ratios ($q$) from optical/infrared 
measurements and pre-main sequence stellar evolution models (see the references 
in Table \ref{table:pairs}).  However, the stellar pairs in this sample have 
wide enough projected separations that they are expected to have prohibitively 
long orbital periods and exhibit little apparent motion on reasonable time 
baselines for observations: therefore, we generally do not have any direct 
knowledge about their true orbital separations ($a$) or eccentricities ($e$).

Nevertheless, we can construct a probabilistic model of $R_t$ using only the 
{\it projected} physical separation of any pair, $a_p$ (based on the projected 
angular separation, $\rho$, and assumed distance, $d$).  Following 
\citet{torres99}, the ratio of the semimajor axis to the projected physical 
separation is
\begin{equation}
\mathcal{F} \equiv \frac{a}{a_p} = \frac{1-e^2}{1+e\cos{\nu}}\sqrt{1-\sin^2{(\omega+\nu)}\sin^2{i}},
\end{equation}
where $e$ is the eccentricity, $\nu$ is the true anomaly, $\omega$ is the 
longitude of periastron, and $i$ is the orbital inclination relative to the 
observer (note that the ratio $\mathcal{F}$ is exact for binaries; we expect 
only modest deviations from it for well-separated hierarchical pairs).  Since 
we do not know \{$\omega$, $i$, $e$, $\nu$\} for any individual stellar pair, 
we have to construct a probability distribution for the true-to-projected 
separation ratios, $\mathcal{P}(\mathcal{F})$, using a Monte Carlo approach.  
To accomplish that, we assume that stellar pairs are not observed at any 
preferential orbital location and adopt uniform distributions for the orbital 
phase (or mean anomaly) and longitude of periastron ($\omega$).  The assumption 
of random viewing geometries suggests that the distribution of orbital 
inclinations ($i$) has a $\sin{i}$ dependence.  However, inferring an 
appropriate functional form for the eccentricity ($e$) distribution (and by 
extension the distribution of true anomalies, $\nu$) is more challenging.

\begin{figure}[t]
\epsscale{1.1}
\plottwo{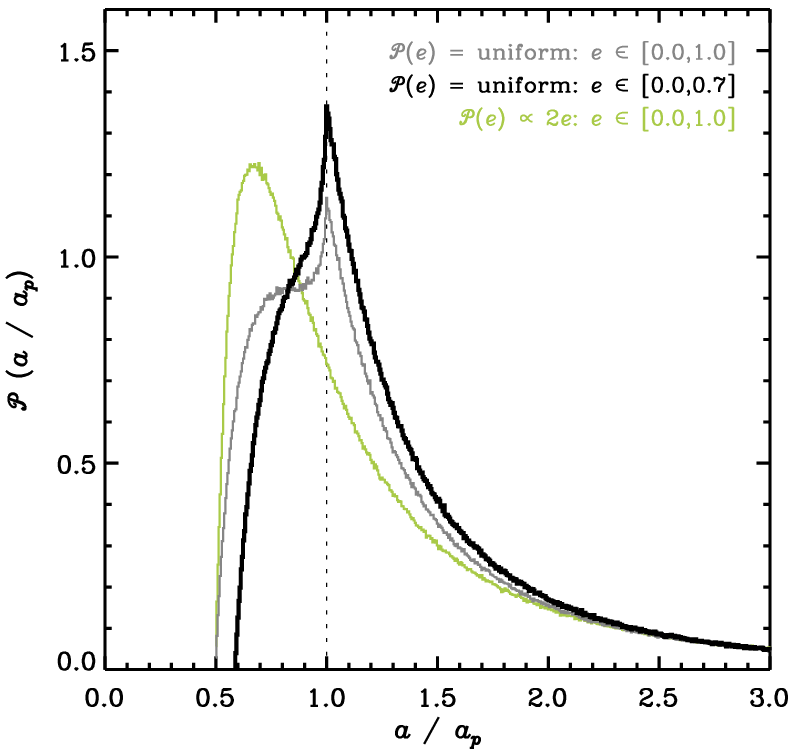}{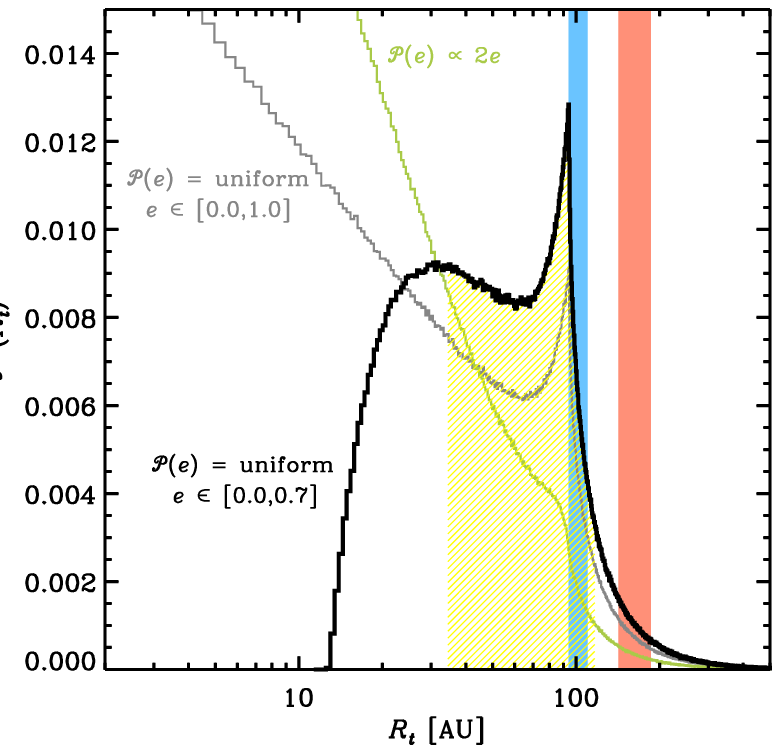}
\figcaption{($a$) Probability distributions for the ratio of true to projected 
separations, $\mathcal{P}(\mathcal{F})$ where $\mathcal{F} \equiv a/a_p$, using 
three different underlying eccentricity distributions.  ($b$) An example 
probability distribution for the disk radii ($R_t$) expected in the HK Tau 
binary based on the tidal interaction models of \citet{pichardo05} for the same 
three assumed eccentricity distributions.  The blue and red vertical bars 
represent the best-fit disk radii (and 1\,$\sigma$ uncertainties) for HK Tau A 
and B, respectively, measured directly from the SMA visibilities in \S 4.  The 
yellow filled area marks the region containing 68\%\ of the probability for the 
$\mathcal{P}(R_t)$ with the favored eccentricity distribution (black curve; see 
text), used to determine the error bars in Figure \ref{fig:trunc_RtRd}.  
\label{fig:trunc_exs}}
\end{figure}

There is little empirical information available to constrain the eccentricity 
distribution for the pre-main sequence binary population.  Pairs with short 
orbital periods have low eccentricities ($e < 0.1$), due to the rapid tidal 
circularization of their orbits \citep{zahn77,zahn89,melo01}.  At longer 
periods, the eccentricity distribution appears relatively uniform in the range 
$e \approx 0.1$-0.9 \citep{mathieu94}.  It is not clear if the apparent dearth 
of young stellar pairs with extreme (circular or parabolic) eccentricities is 
real or an artifact of low-number statistics and selection effects.  Based on 
their samples of main sequence field binaries, \citet{dm91} and 
\citet{tokovinin98} suggested an increasing eccentricity distribution, where 
$\mathcal{P}(e) \propto 2e$ \citep[see also][]{ambartsumian37}.  However, the 
recent comprehensive survey of such systems by \citet{raghavan10} instead 
suggests that a flat eccentricity distribution is preferable for orbital 
periods of $\sim$10-10$^5$ days.  Similar results are also noted for very low 
mass binaries \citep{dupuy11}.  Based on these more recent studies, we assume 
that the eccentricity distribution is uniform.  For reference, Figure 
\ref{fig:trunc_exs}$a$ illustrates the influence of different forms for 
$\mathcal{P}(e)$ on the shape of $\mathcal{P}(\mathcal{F})$. 

Having established the infrastructure to derive a probabilistic model of the 
true orbital separation ($a$) given the observed projected separation ($a_p$), 
we can build on that to determine the distribution of tidal truncation radii, 
$\mathcal{P}(R_t)$, for a given stellar pair.  Because of its relative 
simplicity, we adopt the semi-analytic approximations for truncated disk sizes 
based on the analysis of stable invariant loops by \citet{pichardo05}.  Using 
their formulation, the tidal radius is
\begin{equation}
R_t \approx 0.337 \left[\frac{(1-e)^{1.20} \,\, \varphi^{2/3} \,\, \mu^{0.07}}{0.6 \, \varphi^{2/3}+\ln{(1+\varphi^{1/3})}}\right] \mathcal{F} \,\, a_p \, ,
\end{equation}
where $\mu = q/(1+q)$ is the mass fraction of the stellar pair and $\varphi$ is 
the mass ratio of the host star for which $R_t$ is being calculated relative to 
its companion.  For example, the truncation radius of the disk around the 
primary star is calculated with $\varphi = M_p/M_s = 1/q$, whereas the $R_t$ 
for the disk around the secondary is determined by setting $\varphi = M_s/M_p = 
q$.  For any pair in a multiple system, there is a direct measurement of $a_p$ 
(or rather $\rho$, and an assumed distance) and an estimate of $q$ (see Table 
\ref{table:pairs}).  Fixing these quantities, we constructed the probability 
distribution $\mathcal{P}(R_t)$ that a component of the pair hosts a disk with 
a tidally truncated radius $R_t$ using a Monte Carlo simulation with 
$\sim$10$^7$ realizations of Equation (2), assuming the priors for the 
distributions of the orbital elements \{$\omega$, $i$, $e$, $\nu$\} (and 
therefore $\mathcal{F}$) discussed above.  

As an example, Figure \ref{fig:trunc_exs}$b$ shows $\mathcal{P}(R_t)$ for the 
HK Tau binary, where $a_p = 340$\,AU and $q \approx 1$, for three 
representative assumptions about the underlying eccentricity distribution (note 
that the same $R_t$ distribution applies to both components for this equal-mass 
stellar pair).  The best-fit estimates of the disk radii from our modeling of 
the SMA data (see \S 4) are marked as red (HK Tau A) and blue (HK Tau B) 
vertical bars.  Tidal interaction models predict that the disk sizes have a 
rather steep dependence on the orbital eccentricity (see Equation 2), which 
means that the assumption of an underlying $\mathcal{P}(e)$ that permits or 
favors high eccentricities will lead to the {\it general} prediction of very 
small disk sizes (gray or green curves in Figure \ref{fig:trunc_exs}$b$) and, 
therefore, low millimeter-wave luminosities.  While such eccenticity 
distributions may be relevant for the general population of multiple systems, 
the luminosity-based selection criterion used to build our component-resolved 
SMA sample creates a strong bias that would exclude high-$e$ pairs.  With that 
bias in mind, we favor the use of a truncated eccentricity distribution to 
make comparisons between the measured and predicted disk radii; we assume 
$\mathcal{P}(e)$ is uniform for $e \in [0.0,0.7]$ (black curves in Figure 
\ref{fig:trunc_exs}).  

\begin{figure}[t]
\epsscale{0.55}
\plotone{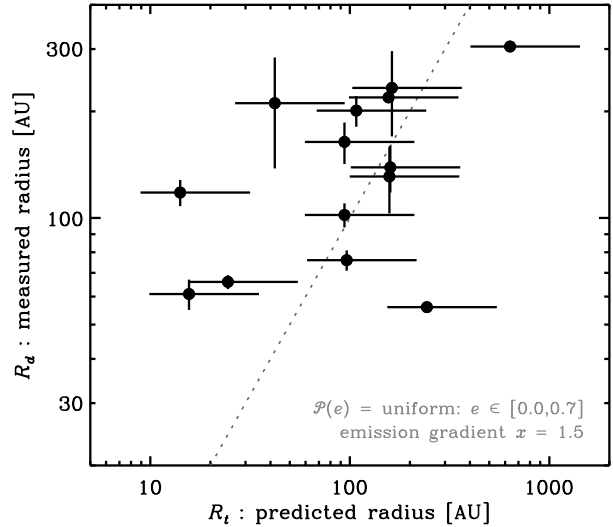}
\figcaption{The measured disk radii (see Tables \ref{table:modelfits} and 
\ref{table:modelfits_unresolved}) compared with the expected disk radii based 
on a probabilistic treatment of the \citet{pichardo05} tidal interaction 
models, for an assumed underlying uniform eccentricity distribution truncated 
at $e = 0.7$ (see text).   \label{fig:trunc_RtRd}}
\end{figure}

Figure \ref{fig:trunc_RtRd} makes a direct comparison between the measurements 
of dust disk sizes ($R_d$) that were determined in \S 4 (see Tables 
\ref{table:modelfits} and \ref{table:modelfits_unresolved}) and the truncation 
radii ($R_t$) predicted by our probabilistic treatment of the 
\citet{pichardo05} models.  The location of the points along the $R_t$ axis 
(abscissae) correspond to the peaks of their $\mathcal{P}(R_t)$ distributions 
(for the assumed $\mathcal{P}(e)$ described above), and their asymmetric error 
bars encapsulate the central 68\%\ of those probability distributions (see the 
shaded yellow region of Figure \ref{fig:trunc_exs}$b$ for an example).  Of the 
14 disks with available $R_d$ measurements, eight have sizes that are in good 
agreement with predictions from the tidal interaction models, two are smaller 
than expected (around GG Tau Aab and DK Tau A), and the remaining four are 
considerably larger.  These decidedly mixed results for the \citet{pichardo05} 
model predictions could be at least partially ameliorated by introducing a 
term that incorporates viscosity into the interaction calculations, which would 
tend to increase $R_t$ due to the viscous spreading of disk material 
\citep[see][]{al94}.  Quantitatively comparable shifts in the measured $R_d$ 
values could be accomodated by permitting a range of emission gradients ($x$) 
in model fits to the observations.  Reconciling the measured disk sizes with 
the predictions from tidal interaction models would require that the disk 
viscosity increases and/or the millimeter-wave emission gradient ($x$) 
decreases as a function of pair separation ($a$).

\section{Discussion}

We have carried out a luminosity and separation limited survey of the 
millimeter-wave dust continuum emission from the disks around individual 
components of 23 young multiple star systems in the $\sim$1-2\,Myr-old 
Taurus-Auriga star formation region.  With a simple morphological model, we 
fitted the continuum visibilities observed with the SMA interferometer to 
determine the luminosity and size of each individual disk in these systems.  
These component-resolved measurements were then coupled with a comprehensive 
database of millimeter-wave luminosities for {\it all} of the multiple systems 
in Taurus (with spectral types F0-M4) to estimate millimeter detection 
frequencies, evaluate the dependence of continuum emission levels on the 
separations and masses of companion stars, and make direct comparisons with 
predictions from tidal interaction models.  

We find that roughly one third (28-37\%) of the individual stars in multiple 
systems harbor disks with dust masses that are large enough to emit detectable 
millimeter continuum radiation (see \S 5.1).  This low incidence rate is 
approximately half that for isolated single stars (62\%), and does not depend 
on the number of stellar companions in the system.  Similar disk frequencies 
have been inferred from accretion and infrared excess signatures
\citep{cieza09,kraus11b}, although without component-resolved diagnostics.  
Given these disk frequencies, it is clear that some {\it external} processes 
driven by the presence of a companion act to hasten the dispersal of 
circumstellar material in multiple systems at a level comparable to any {\it 
internal} disk evolution mechanisms (e.g., photoevaporation, grain growth, 
planet formation, etc.)~on $\sim$1-2\,Myr timescales.  

Some basic demographic properties of stellar pairs in Taurus can provide new 
insights into the mechanics of those external evolution processes.  Building on 
the initial work by \citet{jensen94,jensen96}, we have shown that the 
millimeter-wave luminosity from a pair of stars depends strongly on their 
(projected) separation (see \S 5.2).  We identified substantial changes in the 
millimeter luminosity distributions of pair populations at discrete separations 
of 30 and 300\,AU.  Widely-separated pairs ($a_p > 300$\,AU) have emission 
levels similar to single stars.  Pairs with medium separations ($a_p = 
30$-300\,AU) are typically 5$\times$ fainter, with a lower overall detection 
fraction.  Millimeter emission is rarely detected around pairs with small 
separations ($a_p < 30$\,AU), representing at least another factor of $\sim$5 
reduction in luminosity.  We demonstrated that there is a weak tendency for 
pairs with comparable stellar masses (higher $q$) to have brighter millimeter 
emission, with the effect being considerably stronger for wider pairs.  We 
suggested this is related to a marginal correlation between stellar mass and 
millimeter luminosity, although verifying that tentative trend is a challenge 
due to the systematic uncertainties involved in estimating stellar masses.  

The relationship between millimeter luminosity and pair separation 
(Fig.~\ref{fig:flux_sep}) suggests that the external process relevant for the 
evolution of circumstellar material in multiple systems may be tied to tidal 
stripping from the outer regions of their constituent disks 
\citep[e.g.,][]{al94}.  However, we found mixed results in a comparison of 
individual disk sizes that were measured (from the SMA data; \S 4) and 
predicted from tidal interaction models (see \S 5.3).  Using a probabilistic 
treatment of orbital parameters, we showed that about half of the resolved 
disks in our sample have sizes that are consistent with the truncation 
predictions of the \citet{pichardo05} models.  Most of the remainder have sizes 
that are substantially larger than expected, given the smaller projected 
separations to their companions (Fig.~\ref{fig:trunc_RtRd}).  Analogous 
discrepancies have been noted regarding the observed and predicted inner edges 
of circumbinary rings \citep[e.g.,][]{beust05,nagel10}.  These results hint 
that {\it another} external process also shapes the circumstellar environments 
of young multiples.  Further support for this additional evolutionary process 
is found in the observed distributions of circumstellar dust around individual 
stellar components.  Tidal interaction models predict that disks should survive 
around both components of a pair, with their relative sizes roughly set by $q$ 
(for a given $a$).  However, Fig.~\ref{fig:fratios} (see \S 5.2) demonstrates 
that this is often not the case.  Roughly half of the multiple systems with 
detectable millimeter emission harbor only a single disk, usually around the 
primary component (or the wide tertiary in some hierarchical triples; e.g., HV 
Tau or Haro 6-37).  Moreover, aside from a few exceptional counterexamples (see 
Fig.~\ref{fig:flux_sep}$a$), we find little millimeter-wave evidence for the 
circumbinary disks that should be common around pairs with small separations if 
tidal interactions were the sole external evolution mechanism.  

To be fair, tidal interactions alone may still be able to explain many of these 
observed properties.  The processes of stripping and truncation in pairs where 
the orbital and disk planes are misaligned has not yet been explored in detail 
\citep[although see, e.g.,][]{akeson07,verrier08}, but might be substantially 
enhanced for some configurations.  With limited orbital information, the 
prevalence of such misalignment is not known: but, there is some indication 
from polarization measurements that it is common \citep{jensen04} and a number 
of specific examples have been identified \citep[e.g., HK Tau or HD 
98800;][]{duchene03,andrews10}.  Moreover, it is important to keep in mind that 
the emission we have measured is only capable of probing the trace population 
of dust particles in these disks, and not their dominant mass reservoirs of 
molecular gas.  The total disk masses that would be inferred from this emission 
could be substantially under-estimated \citep[see][]{aw05,aw07}, as could the 
apparent disk sizes \citep[e.g.,][]{hughes08,panic09,andrews12}.  Future 
complementary observations of emission line probes would help better address 
such uncertainties.

Alternatively, the disk properties we observe may be set at very early stages 
by the processes that regulate accretion during multiple star formation.  
Numerical simulations indicate that the ratio of the specific angular momenta 
of the infalling material ($j_{\rm gas}$) to the stellar pair ($j_{\ast}$) is 
fundamental in determining how the gas and dust accreted from a proto-system 
envelope is distributed among individual stellar components 
\citep{bonnell94,bate97}.  If $j_{\rm gas} < j_{\ast}$, most of the infalling 
material will form a disk around the stellar primary.  Conversely, if $j_{\rm 
gas} \ge j_{\ast}$, then a circum-secondary or circumbinary disk will dominate 
\citep[see also][]{ochi05}.  In the many cases with only circum-primary disk 
detections described above, the data are consistent with the former scenario.  
Similar millimeter-wave observations suggest that primary stars preferentially 
harbor more circumstellar material at the even earlier Class 0/I stages of 
protostellar evolution \citep[e.g.,][and references 
therein]{launhardt04,patience08}.

To summarize, there is a body of observational evidence suggesting that (at 
least) two fundamental processes related to the presence of stellar companions 
play significant roles in the evolution of circumstellar material in young 
multiple systems.  The first is associated with the multiple star formation 
process itself, where the fraction of angular momentum associated with 
infalling material relative to that contained in the orbital motion of the 
stellar pair determines how circumstellar material is apportioned to each 
component.  This is likely responsible for the pairs we observe with very large 
primary-to-secondary millimeter luminosity ratios.  The second is a tidal 
interaction process that strips material from any individual disks that survive 
the formation process.  This is thought to be the origin of the millimeter 
luminosity -- pair separation relationship.

At the typical $\sim$1-2\,Myr age of Taurus-Auriga multiples, both of these 
evolution mechanisms have already made their mark on disk properties.  The 
long-term fate of the circumstellar material in these multiple systems now 
rests with the same internal mechanisms that govern the subsequent evolution 
and dissipation of the disks around their isolated (single) counterparts.  The 
formation and evolution of planets from this disk material are mechanisms of 
particular interest in these systems.  The relatively straightforward, albeit 
uncertain, relationship between millimeter-wave luminosities and total disk 
masses enables at least a preliminary assessment of the likelihood of planet 
formation around individual components in multiple systems.  Using the 
conversion advocated by \citet{aw05,aw07} -- which assumes optically thin, 
isothermal dust emission with a mean temperature of 20\,K and (gas+dust) 
opacity of 0.034\,cm$^2$ g$^{-1}$ at 880\,$\mu$m -- $F_d \approx 15$\,mJy 
corresponds to a disk mass of $\sim$1\,M$_{\rm Jup}$, and $F_d \approx 
150$\,mJy represents the standard estimate of the minimum mass of the solar 
nebula, $\sim$0.01\,M$_{\odot}$.  Taking these conversions at face value, we 
would conclude that giant planets are unlikely to form around stars in close 
pairs (Fig.~\ref{fig:flux_sep}) or around the secondary components of most 
pairs with wider separations (Fig.~\ref{fig:fratios}).  However, we would 
likewise infer that the primary components in wider pairs, the wide tertiaries 
in hierarchical triples, and perhaps the population of spectroscopic binaries 
should be just as likely to host giant planets as single stars.  

Direct exoplanet searches of main sequence multiple systems in the field 
confirm these general expectations from the disk survey: planet formation is 
not severely inhibited by the presence of a stellar companion 
\citep[e.g.,][]{patience02,raghavan06,duchene10}, and giant planets are 
preferentially found around primaries, the isolated components of hierarchical 
triples \citep{desidera07,bonavita07,mugrauer09,desidera11}, and perhaps even 
spectroscopic binaries \citep[e.g.,][]{doyle11}.  That corroboration of results 
is a promising sign for planet formation studies in multiple systems, but the 
characterization of the disks around individual stellar components in these 
systems is still in its early stages.  In the near future, we expect that new 
facilities like the Atacama Large Millimeter Array (ALMA) will shift the focus 
to study these individual disks in detail to directly compare their density 
structures and particle growth signatures with the disks around isolated 
(single) stars.  Ultimately, such observations will help determine the impact 
of a nearby stellar neighbor on the planet formation process.

\section{Summary}

We have presented the results of an SMA imaging survey of the millimeter-wave 
(880\,$\mu$m or 1.3\,mm) thermal continuum emission from circumstellar dust in 
23 young multiple systems in the Taurus-Auriga star-forming region.  This 
census was designed to target relatively bright ($>$20\,mJy at 880\,$\mu$m) and 
well-separated ($\rho > 0\farcs3$) systems with primary spectral types between 
F0 and M4.  We employed simple morphological models of the SMA visibilities to 
measure component-resolved millimeter luminosities and disk sizes whenever 
possible.  Those results were considered together with a comprehensive 
literature compilation of the millimeter luminosities from {\it all} Taurus 
multiples (in this spectral type range) to better analyze how the presence of a 
stellar compaion affects basic disk properties.  Our primary conclusions are 
the following:

\begin{enumerate}
\item The millimeter detection frequency for individual stars in multiple 
systems is approximately 1/3 (28-40\%), about half that for single stars 
(62\%), and independent of the number of companions.  These relative incidence 
rates suggest that the presence of a stellar companion plays a substantial {\it 
external} role in the early development and evolution of circumstellar 
material, at a level comparable to the standard {\it internal} disk evolution 
mechanisms that can operate in isolation (e.g., photoevaporation, particle 
growth, planet formation, etc.).

\item The millimeter luminosity from a pair of stars depends strongly on their 
projected separation ($a_p$), such that closer pairs are substantially 
fainter.  We find natural breaks in the luminosity--separation plane at $a_p 
\approx 30$ and 300\,AU.  The luminosity distribution of wide pairs ($a_p > 
300$\,AU) is indistinguishable from that of single stars.  Pairs with medium 
separations ($a_p = 30$-300\,AU) are $5\times$ fainter, and the very few close 
pairs ($a_p < 30$\,AU) that we detect are $5\times$ fainter yet --
although a few bright circum{\it binary} disks represent notable exceptions.  

\item There is no clear relationship between the millimeter luminosity from a 
pair and its stellar mass ratio ($q$) in general, but wide pairs with higher 
$q$ tend to be brighter.  We show that this latter behavior is produced by a 
marginal correlation between millimeter luminosities and stellar masses (both 
summed among pairs and for individual stars).  However, the significance of 
this trend is questionable: the scatter is large, and there are substantial 
systematic uncertainties in estimating individual stellar masses that are not 
considered here.  In nearly all cases, the primary component of a binary pair 
or the wide tertiary of a hierarchical triple harbors the disk material that 
dominates the millimeter luminosity of the system (higher-order systems 
show a range of behavior, depending on the hierarchical nature of their 
pairings).  

\item We find mixed results from a direct comparison of the disk sizes measured 
from the data and predicted from tidal interaction models, based on a 
probabilistic treatment of the orbital parameters for each system.  Of the 15 
resolved disks in our sample, the radii expected from the models described by 
\citet{pichardo05} are commensurate with the observed radii for eight disks; 
two others are found to be too small, and the remaining five are notably larger 
than would be expected given the relatively small (projected) distances to 
their companions.  

\item These millimeter-wave observations suggest that at least two external 
mechanisms contribute to the evolution of circumstellar material in young 
multiple star systems.  Star-disk tidal interactions strip material from the 
outer regions of individual disks, a process responsible for the strong 
dependence of the millimeter luminosity on the projected separation to a 
companion.  We are lead to infer that accretion during the multiple star 
formation process itself also plays a substantial role in apportioning disk 
material, in many cases setting the initial disk masses such that primary stars 
harbor substantially more circumstellar material than their companions.  The 
long-term prospects for (giant) planet formation in multiple systems are poor 
for stars in close pairings and around secondaries, but should be comparable to 
those around single stars for primaries, wide tertiaries in hierarchical 
triples, and perhaps around both components of very close spectroscopic 
binaries.  
\end{enumerate}

\acknowledgments We are grateful to Trent Dupuy for his assistance with the 
probabilistic treatment of projected orbits, to Joanna Brown for kindly 
providing some supplementary observing time on the FQ Tau binary, and to an 
anonymous referee for helpful suggestions on clarifying the draft manuscript.  
The Submillimeter Array (SMA) is a joint project between the Smithsonian 
Astrophysical Observatory and the Academia Sinica Institute of Astronomy and 
Astrophysics and is funded by the Smithsonian Institution and the Academia 
Sinica.

\input{table1.tex}
\input{table2.tex}
\input{table3.tex}
\input{table4.tex}
\input{table5.tex}
\input{table6.tex}
\input{table7.tex}
\input{table8.tex}
\input{table9.tex}

\clearpage

\bibliography{references}

\end{document}

%% file: table1.tex
\begin{deluxetable}{lcccc}
\tablecolumns{5}
\tablewidth{0pc}
\tabletypesize{\footnotesize}
\tablecaption{Selected Sample of Taurus Multiples\label{table:sample}}
\tablehead{
\colhead{System} & \colhead{} & \colhead{$N_{\ast}$} & \colhead{} & \colhead{$N_{\rm pair}$}}  
\startdata
MHO 1/MHO 2            & \phantom{blank} & 3 & \phantom{blank} & 2 \\
DD Tau                 & \phantom{blank} & 2 & \phantom{blank} & 1 \\
FQ Tau                 & \phantom{blank} & 2 & \phantom{blank} & 1 \\
T Tau                  & \phantom{blank} & 3 & \phantom{blank} & 2 \\
{\it FS Tau/Haro 6-5B} & \phantom{blank} & 3 & \phantom{blank} & 2 \\ 
FV Tau                 & \phantom{blank} & 4 & \phantom{blank} & 3 \\
DI Tau/DH Tau          & \phantom{blank} & 4 & \phantom{blank} & 3 \\
UX Tau                 & \phantom{blank} & 4 & \phantom{blank} & 4 \\
\enddata
\tablecomments{Col.~(1): System name.  Italicized systems have not been 
re-observed with the SMA (see \S 2).  Col~(2): Number of stars in system.  
Col.~(3): Number of ``pairs" in system (see \S 2).}  
\end{deluxetable}

%% file: table2.tex
\begin{deluxetable}{llcc}
\tablecolumns{4}
\tablewidth{0pc}
\tabletypesize{\footnotesize}
\tablecaption{SMA Observing Log\label{table:log}}
\tablehead{
\colhead{UT Date} & \colhead{Targets} & \colhead{Config.} & \colhead{$\nu$(LO)} \\
\colhead{}        & \colhead{}        & \colhead{}        & \colhead{[GHz]}} 
\startdata
2005 Dec 17       & DH/DI Tau                       & C & 340.755 \\
2009 Oct 24       & DK Tau                          & C & 340.767 \\
2010 Feb 11       & DK Tau                          & V & 343.277 \\
2010 Feb 18       & UZ Tau                          & V & 341.536 \\
2010 Feb 19       & HK Tau                          & V & 342.289 \\
2010 Mar 2        & UZ Tau                          & V & 340.427 \\
2010 Nov 10       & IT Tau, FX Tau, HK Tau,         & C & 340.755 \\
                  & V710 Tau, HN Tau, CIDA-9        &   &         \\
\enddata
\tablecomments{Col.~(1): UT date of observations.  Col.~(2): Multiple systems 
observed.  Col.~(3): SMA antenna configuration (see \S 3).  Col.~(4): LO 
frequency setting.  Note that the observations below the horizontal line 
focused on multiple systems that are {\it not} part of our sample: they were 
observed because there were no previous millimeter-wave measurements of these 
systems available in the literature (see \S 5.1).} 
\end{deluxetable}

%% file: table3.tex
\begin{deluxetable}{lcccc}
\tablecolumns{5}
\tablewidth{0pc}
\tabletypesize{\footnotesize}
\tablecaption{Properties of Synthesized Continuum Maps\label{table:maps}}
\tablehead{
\colhead{System} & \colhead{$\lambda$} & \colhead{RMS noise}         & \colhead{beam size} & \colhead{beam PA} \\
\colhead{}       & \colhead{[$\mu$m]}  & \colhead{[mJy beam$^{-1}$]} & \colhead{[\arcsec]}           & \colhead{[\degr]}         \\
\colhead{(1)}    & \colhead{(2)}       & \colhead{(3)}               & \colhead{(4)}                 & \colhead{(5)}}
\startdata
MHO 1/2   & 1335 & 2.1 & $2.8\times2.2$ & 95  \\
DD Tau    & 880  & 1.9 & $0.9\times0.7$ & 79  \\
FQ Tau    & 880  & 1.2 & $0.9\times0.7$ & 77  \\
T Tau     & 880  & 3.6 & $1.0\times0.8$ & 77  \\
FV Tau    & 880  & 1.1 & $1.0\times0.7$ & 79  \\
DI/DH Tau & 880  & 3.3 & $2.6\times2.1$ & 151 \\
UX Tau    & 880  & 0.9 & $0.6\times0.4$ & 33  \\
FX Tau    & 880  & 1.6 & $1.1\times0.8$ & 82  \\
DK Tau    & 880  & 0.7 & $0.5\times0.5$ & 0   \\
HK Tau    & 880  & 1.4 & $0.7\times0.6$ & 65  \\
V710 Tau  & 880  & 3.2 & $2.2\times2.0$ & 94  \\
GG Tau    & 1335 & 2.2 & $3.2\times2.4$ & 32  \\
UZ Tau    & 880  & 0.7 & $0.4\times0.3$ & 51  \\
JH 112    & 880  & 1.9 & $2.0\times1.8$ & 57  \\
GI/GK Tau & 1335 & 0.7 & $3.6\times2.6$ & 68  \\
HN Tau    & 880  & 2.1 & $2.2\times2.0$ & 96  \\
IT Tau    & 880  & 1.4 & $2.1\times1.9$ & 78  \\
HP Tau    & 1335 & 1.2 & $3.9\times2.9$ & 102 \\
HV Tau    & 880  & 2.3 & $2.0\times1.8$ & 61  \\
Haro 6-37 & 880  & 5.0 & $2.1\times2.0$ & 136 \\
UY Aur    & 880  & 1.5 & $0.9\times0.6$ & 91  \\
CIDA-9    & 880  & 2.4 & $2.2\times2.0$ & 98  \\
RW Aur    & 880  & 3.6 & $2.1\times2.0$ & 159 \\
\enddata
\tablecomments{Col.~(1): System name.  Col.~(2): Effective observing 
wavelength.  Col.~(3): RMS noise level (1\,$\sigma$) in the naturally weighted 
synthesis image (see Figures \ref{fig:images1}-\ref{fig:images3}).  Col.~(4) 
\& (5): Beam dimensions and major axis position angle.}
\end{deluxetable}

%% file: table4.tex
\begin{deluxetable}{lccrrrr}
\tablecolumns{7}
\tablewidth{0pc}
\tabletypesize{\footnotesize}
\tablecaption{Resolved Disk Properties\label{table:modelfits}}
\tablehead{
\colhead{Name} & \multicolumn{2}{c}{Disk Center} & \colhead{$F_d$} & \colhead{$R_d$} & \colhead{$i_d$}   & \colhead{PA$_d$}  \\
\colhead{}     & \colhead{$\alpha$ [J2000]} & \colhead{$\delta$ [J2000]} & \colhead{[mJy]} & \colhead{[AU]}  & \colhead{[\degr]} & \colhead{[\degr]} \\
\colhead{(1)} & \colhead{(2)} & \colhead{(3)} & \colhead{(4)} & \colhead{(5)} & \colhead{(6)} & \colhead{(7)}
}
\startdata
MHO 1          & 04 14 26.28 & +28 06 02.90 & ($242\pm3$) & $131\pm28$ & $32\pm15$  & $84\pm28$  \\
T Tau N        & 04 21 59.44 & +19 32 06.31 & $304\pm3$   & $64\pm3$   & $6\pm21$   & $90\pm36$  \\
UX Tau A       & 04 30 04.00 & +18 13 49.25 & $141\pm1$   & $57\pm1$   & $34\pm2$   & $174\pm4$  \\  
FX Tau A       & 04 30 29.64 & +24 26 44.77 & $24\pm3$    & $219\pm57$ & [45]       & [90]       \\ 
DK Tau A       & 04 30 44.25 & +26 01 24.52 & $57\pm1$    & $76\pm5$   & $27\pm9$   & $115\pm12$ \\ 
HK Tau A       & 04 31 50.57 & +24 24 17.60 & $81\pm2$    & $102\pm8$  & $57\pm4$   & $118\pm5$  \\
HK Tau B       & 04 31 50.60 & +24 24 15.27 & $49\pm2$    & $164\pm22$ & [84]       & [40]       \\
GG Tau Aab     & 04 32 30.36 & +17 31 40.03 & ($557\pm3$) & $305\pm5$  & [37]       & [7]        \\
\enddata
\tablecomments{Col.~(1): Stellar component name.  Col.~(2) and (3): Adopted 
disk center coordinates.  Col.~(4): Flux density.  Values in parenthesis 
correspond to measurements at a wavelength of 1.3\,mm: all others are at 
880\,$\mu$m.  Col.~(5): Disk radius.  Col.~(6): Disk inclination angle, where 
0\degr\ is face-on.  Col.~(7): Position angle (E of N) of the disk major axis 
projected on the sky.  Values in square brackets were fixed in the modeling 
(see \S 4).}  
\end{deluxetable}

%% file: table5.tex
\begin{deluxetable}{lccr|lccr}
\rotate
\tablecolumns{8}
\tablewidth{0pc}
\tabletypesize{\footnotesize}
\tablecaption{Unresolved Disk Properties\label{table:modelfits_unresolved}}
\tablehead{
\colhead{Name} & \multicolumn{2}{c}{Disk Center} & \colhead{$F_d$} & \colhead{Name} & \multicolumn{2}{c}{Disk Center} & \colhead{$F_d$} \\
\colhead{}     & \colhead{$\alpha$ [J2000]} & \colhead{$\delta$ [J2000]} & \colhead{[mJy]} & \colhead{}     & \colhead{$\alpha$ [J2000]} & \colhead{$\delta$ [J2000]} & \colhead{[mJy]} 
}
\startdata
MHO 2 AB   \phantom{sss} & 04 14 26.47 & +28 05 59.95 & ($138\pm1$) \phantom{s} & \phantom{s} GG Tau Bb     & 04 32 30.51 & +17 31 28.68 & \phantom{s} ($<7$)     \phantom{s} \\
DD Tau A   \phantom{sss} & 04 18 31.13 & +28 16 28.82 & $24\pm3$    \phantom{s} & \phantom{s} {\it FZ Tau}        & 04 32 31.76 & +24 20 03.10 & \phantom{s} $29\pm7$   \phantom{s} \\
DD Tau B   \phantom{sss} & 04 18 31.13 & +28 16 28.26 & $9\pm3$     \phantom{s} & \phantom{s} {\it FY Tau}        & 04 32 30.58 & +24 19 57.29 & \phantom{s} $<27$      \phantom{s} \\
FQ Tau A   \phantom{sss} & 04 19 12.81 & +28 29 32.75 & $8\pm2$     \phantom{s} & \phantom{s} JH 112 Aa     & 04 32 49.12 & +22 53 02.52 & \phantom{s} $10\pm3$   \phantom{s} \\
FQ Tau B   \phantom{sss} & 04 19 12.79 & +28 29 33.44 & $<5$        \phantom{s} & \phantom{s} JH 112 Ab     & 04 32 49.22 & +22 53 03.04 & \phantom{s} $5\pm3$    \phantom{s} \\
T Tau Sab  \phantom{sss} & 04 21 59.44 & +19 32 05.61 & $<10$       \phantom{s} & \phantom{s} JH 112 B      & 04 32 49.27 & +22 53 07.89 & \phantom{s} $<6$       \phantom{s} \\
{\it FS Tau AB}  \phantom{sss} & 04 22 02.18 & +26 57 30.49 & $49\pm6$    \phantom{s} & \phantom{s} {\it V807 Tau AB}   & 04 33 06.64 & +24 09 54.90 & \phantom{s} $20\pm3$   \phantom{s} \\
{\it Haro 6-5B}  \phantom{sss} & 04 22 00.70 & +26 57 32.50 & ($134\pm6$) \phantom{s} & \phantom{s} {\it GH Tau AB}     & 04 33 06.22 & +24 09 33.99 & \phantom{s} $15\pm3$   \phantom{s} \\
\enddata
\tablecomments{Col.~(1): Stellar component name.  Col.~(2) and (3): Disk center coordinates.
Col.~(4): 880\,$\mu$m flux density.  Values in parenthesis correspond to 1.3\,mm.  The sources in 
italics have $F_d$ values from single-dish photometry \citep{aw05}.}
\end{deluxetable}

%% file: table6.tex
\begin{deluxetable}{lllcccrr}
\tablecolumns{8}
\tablewidth{0pc}
\tabletypesize{\footnotesize}
\tablecaption{Properties of Stellar Pairs in Taurus\label{table:pairs}}
\tablehead{
\colhead{System}    & \multicolumn{2}{c}{Pair}                & \colhead{$\rho$}    & \colhead{$q$} & \colhead{$M_p$}         & \colhead{$F_{\rm pair}$} & \colhead{Refs} \\
\colhead{}          & \colhead{Primary} & \colhead{Secondary} & \colhead{[\arcsec]} & \colhead{}    & \colhead{[M$_{\odot}$]} & \colhead{[mJy]}          & \colhead{}     \\
\colhead{(1)}       & \colhead{(2)}     & \colhead{(3)}       & \colhead{(4)}       & \colhead{(5)} & \colhead{(6)}           & \colhead{(7)}            & \colhead{(8)}}
\startdata
J04080782+2807280   & J04080782 A       & J04080782 B         & 0.044               & 0.13          & 0.30                    & ($<8$)                   & 1, 1, A                                                     \\
V773 Tau            & V773 Tau Aa       & V773 Tau Ab         & 0.0028              & 0.86          & 1.54                    & $<9$                     & 2, 2, B\tablenotemark{a}                                  \\
                    & V773 Tau Aab      & V773 Tau B          & 0.14                & 0.29          & 2.87                    & $<9$                     & 3, 3\tablenotemark{b},B\tablenotemark{a}                  \\
                    & V773 Tau Aab      & V773 Tau C          & 0.21                & \nodata       & 2.87                    & $<9$                     & 3, 3\tablenotemark{b},B\tablenotemark{a}                  \\
                    & V773 Tau B        & V773 Tau C          & 0.20                & \nodata       & 1.33                    & $<9$                     & 3\tablenotemark{c},3\tablenotemark{b},B\tablenotemark{a} \\
                    & V773 Tau ABC      & J0414118+2811535    & 23.38               & $\sim$0.03    & $>4.20$                 & ($<9$)                   & 4, 1\tablenotemark{b},A                                    \\
Anon 1              & Anon 1 A          & Anon 1 B            & 0.015               & 0.68          & 0.64                    & $<8$                     & 1, 1, B                                                     \\
MHO 1/2             & MHO 1             & MHO 2 AB            & 3.93                & 1.00          & 0.44                    & ($380\pm3$)              & 4, 1, A                                                     \\
\enddata
\end{deluxetable}

%% file: table7.tex
\begin{deluxetable}{lcrc|lcrc}
\tablecolumns{8}
\tablewidth{0pc}
\tabletypesize{\footnotesize}
\tablecaption{Single Stars Observed at Millimeter Wavelengths\label{table:singles}}
\tablehead{
\colhead{Name} & \colhead{$M_{\ast}$}    & \colhead{$F_d$} & \colhead{Ref} & \colhead{Name} & \colhead{$M_{\ast}$}    & \colhead{$F_d$} & \colhead{Ref} \\
\colhead{}     & \colhead{[M$_{\odot}$]} & \colhead{[mJy]} & \colhead{}    & \colhead{}     & \colhead{[M$_{\odot}$]} & \colhead{[mJy]} & \colhead{} \\
\colhead{(1)} & \colhead{(2)} & \colhead{(3)} & \colhead{(4)} & \colhead{(1)} & \colhead{(2)} & \colhead{(3)} & \colhead{(4)}}
\startdata
LkCa 1           & 0.27 & $<8$         & 1, A \phantom{a} & \phantom{a} L1551-55        & 0.72 & $<5$         & 1, A \\
IRAS 04108+2910  & 0.30 & ($<20$)      & 1, B \phantom{a} & \phantom{a} V830 Tau        & 0.72 & $<6$         & 1, A \\
FM Tau           & 0.64 & $32\pm8$     & 1, A \phantom{a} & \phantom{a} IRAS 04301+2608 & 0.64 & $18\pm6$     & 1, A \\
CW Tau           & 0.94 & $66\pm6$     & 1, A \phantom{a} & \phantom{a} DL Tau          & 0.72 & $440\pm40$   & 1, A \\
FP Tau           & 0.27 & ($<9$)       & 1, B \phantom{a} & \phantom{a} DM Tau          & 0.57 & $210\pm1$    & 1, E \\
CX Tau           & 0.45 & $25\pm6$     & 1, A \phantom{a} & \phantom{a} CI Tau          & 0.72 & $324\pm6$    & 1, A \\
LkCa 4           & 0.72 & $<4$         & 1, A \phantom{a} & \phantom{a} JH 108          & 0.57 & $<7$         & 1, A \\
CY Tau           & 0.57 & $140\pm5$    & 1, A \phantom{a} & \phantom{a} AA Tau          & 0.72 & $144\pm5$    & 1, A \\
\enddata
\tablecomments{Col.~(1): Star name.  Col.~(2): Estimated stellar mass.  
Col.~(3): 880\,$\mu$m flux density or upper limit.  Values in parenthesis were 
measured instead at 1.3\,mm: the analysis in \S 5 assumes a conservative 
scaling by a factor $(1.3/0.88)^2 \approx 2.2$.  Col.~(4): References for 
$M_{\ast}$ and $F_d$: [1] = \citet{kraus11}, and [A] = \citet{aw05}, [B] = 
\citet{schaefer09}, [C] = \citet{motte01}, [D] = \citet{mannings94}, and [E] = 
\citet{andrews11}.} 
\end{deluxetable}

%% file: table8.tex
\begin{deluxetable}{lccr}
\tablecolumns{4}
\tablewidth{0pc}
\tabletypesize{\footnotesize}
\tablecaption{Millimeter Detection Statistics\label{table:fmm}}
\tablehead{
\colhead{Type} & \colhead{$N_{\rm det}$} & \colhead{$N_{\rm tot}$} & \colhead{$f_{\rm mm}$} }
\startdata
singles       & 32    & 52  & $62\pm11$\%\    \\
binaries      & 26-34 & 96  & 27-$35\pm5$\%\  \\
triples       & 11-15 & 39  & 28-$38\pm10$\%\ \\
higher order  & 13-18 & 44  & 30-$41\pm10$\%\ \\
all multiples & 50-67 & 179 & 28-$37\pm5$\%\  
\enddata
\tablecomments{Col.~(1): Type of stellar system.  Col.~(2): Number of 
individual stars with detected millimeter emission.  Col.~(3): Total number of 
stars of that category.  Col.~(4): Millimeter detection fraction.}
\end{deluxetable}

%% file: table9.tex
\begin{deluxetable}{llcc}
\tablecolumns{4}
\tablewidth{0pc}
\tabletypesize{\footnotesize}
\tablecaption{Two-Sample Tests Comparing Millimeter Luminosity Distributions\label{table:surv_tests}}
\tablehead{
\colhead{Group 1} & \colhead{Group 2} & \colhead{$\mathcal{P}$(survival)} & \colhead{$\mathcal{P}$(K-S)} }
\startdata
single stars               & wide pairs ($>$300\,AU)   & 0.20-0.40  & 0.37  \\
wide pairs ($>$300\,AU)    & medium pairs (30-300\,AU) & $<$0.00006 & 0.017 \\
medium pairs (30-300\,AU)  & small pairs ($<$30\,AU)   & $<$0.00005 & 0.047 \\
\hline
high mass ratios ($q>0.5$) & low mass ratios ($q<0.5$) & $<$0.04    & 0.48  \\
wide pairs with $q>0.5$    & wide pairs with $q<0.5$   & $<$0.00001 & 0.77 \\
medium pairs with $q>0.5$  & medium pairs with $q<0.5$ & $<$0.08    & 0.10 \\
\hline
single stars		   & primaries in pairs        & 0.41-0.98  & \nodata \\
primaries in pairs         & secondaries in pairs      & $<$0.008   & \nodata \\
\enddata
\tablecomments{Cols.~(1) \& (2): The sub-populations being compared.  Col.~(3): 
The probability that these two groups are drawn from the same parent 
distributions.  The listed values correspond to the range or more conservative 
limit from the standard two-sample tests used in survival analysis 
(Peto-Prentice, Logrank, and Gehan tests).  Col.~(4): Same as Col.~(3), but for 
the special case that includes only those pairs with millimeter-wave detections 
in each sub-population, derived from the two-sided Kolmogorov-Smirnov test.}
\end{deluxetable}

%% file: ms.bbl
\begin{thebibliography}{99}
\expandafter\ifx\csname natexlab\endcsname\relax\def\natexlab#1{#1}\fi

\bibitem[{{Abt} \& {Levy}(1976)}]{abt76}
{Abt}, H.~A., \& {Levy}, S.~G. 1976, \apjs, 30, 273

\bibitem[{{Akeson} {et~al.}(1998){Akeson}, {Koerner}, \& {Jensen}}]{akeson98}
{Akeson}, R.~L., {Koerner}, D.~W., \& {Jensen}, E.~L.~N. 1998, \apj, 505, 358

\bibitem[{{Akeson} {et~al.}(2007){Akeson}, {Rice}, {Boden}, {Sargent},
  {Carpenter}, \& {Bryden}}]{akeson07}
{Akeson}, R.~L., {Rice}, W.~K.~M., {Boden}, A.~F., {et~al.} 2007, \apj, 670,
  1240

\bibitem[{{Ambartsumian}(1937)}]{ambartsumian37}
{Ambartsumian}, V.~A. 1937, AZh, 14, 207

\bibitem[{{Andrews} {et~al.}(2010){Andrews}, {Czekala}, {Wilner}, {Espaillat},
  {Dullemond}, \& {Hughes}}]{andrews10}
{Andrews}, S.~M., {Czekala}, I., {Wilner}, D.~J., {et~al.} 2010, \apj, 710, 462

\bibitem[{{Andrews} \& {Williams}(2005)}]{aw05}
{Andrews}, S.~M., \& {Williams}, J.~P. 2005, \apj, 631, 1134

\bibitem[{{Andrews} \& {Williams}(2007{\natexlab{a}})}]{aw07}
---. 2007{\natexlab{a}}, \apj, 671, 1800

\bibitem[{{Andrews} \& {Williams}(2007{\natexlab{b}})}]{aw07b}
---. 2007{\natexlab{b}}, \apj, 659, 705

\bibitem[{{Andrews} {et~al.}(2011){Andrews}, {Wilner}, {Espaillat}, {Hughes},
  {Dullemond}, {McClure}, {Qi}, \& {Brown}}]{andrews11}
{Andrews}, S.~M., {Wilner}, D.~J., {Espaillat}, C., {et~al.} 2011, \apj, 732,
  42

\bibitem[{{Andrews} {et~al.}(2012){Andrews}, {Wilner}, {Hughes}, {Qi},
  {Rosenfeld}, {{\"O}berg}, {Birnstiel}, {Espaillat}, {Cieza}, {Williams},
  {Lin}, \& {Ho}}]{andrews12}
{Andrews}, S.~M., {Wilner}, D.~J., {Hughes}, A.~M., {et~al.} 2012, \apj, 744,
  162

\bibitem[{{Artymowicz} \& {Lubow}(1994)}]{al94}
{Artymowicz}, P., \& {Lubow}, S.~H. 1994, \apj, 421, 651

\bibitem[{{Baraffe} {et~al.}(1998){Baraffe}, {Chabrier}, {Allard}, \&
  {Hauschildt}}]{baraffe98}
{Baraffe}, I., {Chabrier}, G., {Allard}, F., \& {Hauschildt}, P.~H. 1998, \aap,
  337, 403

\bibitem[{{Bate} \& {Bonnell}(1997)}]{bate97}
{Bate}, M.~R., \& {Bonnell}, I.~A. 1997, \mnras, 285, 33

\bibitem[{{Beckwith} {et~al.}(1990){Beckwith}, {Sargent}, {Chini}, \&
  {Guesten}}]{beckwith90}
{Beckwith}, S.~V.~W., {Sargent}, A.~I., {Chini}, R.~S., \& {Guesten}, R. 1990,
  \aj, 99, 924

\bibitem[{{Beust} \& {Dutrey}(2005)}]{beust05}
{Beust}, H., \& {Dutrey}, A. 2005, \aap, 439, 585

\bibitem[{{Bieging} {et~al.}(1984){Bieging}, {Cohen}, \&
  {Schwartz}}]{bieging84}
{Bieging}, J.~H., {Cohen}, M., \& {Schwartz}, P.~R. 1984, \apj, 282, 699

\bibitem[{{Boden} {et~al.}(2007){Boden}, {Torres}, {Sargent}, {Akeson},
  {Carpenter}, {Boboltz}, {Massi}, {Ghez}, {Latham}, {Johnston}, {Menten}, \&
  {Ros}}]{boden07}
{Boden}, A.~F., {Torres}, G., {Sargent}, A.~I., {et~al.} 2007, \apj, 670, 1214

\bibitem[{{Bonavita} \& {Desidera}(2007)}]{bonavita07}
{Bonavita}, M., \& {Desidera}, S. 2007, \aap, 468, 721

\bibitem[{{Bonnell} \& {Bate}(1994)}]{bonnell94}
{Bonnell}, I.~A., \& {Bate}, M.~R. 1994, \mnras, 269, L45

\bibitem[{{Carrasco-Gonz{\'a}lez} {et~al.}(2009){Carrasco-Gonz{\'a}lez},
  {Rodr{\'{\i}}guez}, {Anglada}, \& {Curiel}}]{carrasco-gonzalez09}
{Carrasco-Gonz{\'a}lez}, C., {Rodr{\'{\i}}guez}, L.~F., {Anglada}, G., \&
  {Curiel}, S. 2009, \apjl, 693, L86

\bibitem[{{Cieza} {et~al.}(2009){Cieza}, {Padgett}, {Allen}, {McCabe},
  {Brooke}, {Carey}, {Chapman}, {Fukagawa}, {Huard}, {Noriga-Crespo},
  {Peterson}, \& {Rebull}}]{cieza09}
{Cieza}, L.~A., {Padgett}, D.~L., {Allen}, L.~E., {et~al.} 2009, \apjl, 696,
  L84

\bibitem[{{Correia} {et~al.}(2006){Correia}, {Zinnecker}, {Ratzka}, \&
  {Sterzik}}]{correia06}
{Correia}, S., {Zinnecker}, H., {Ratzka}, T., \& {Sterzik}, M.~F. 2006, \aap,
  459, 909

\bibitem[{{Desidera} \& {Barbieri}(2007)}]{desidera07}
{Desidera}, S., \& {Barbieri}, M. 2007, \aap, 462, 345

\bibitem[{{Desidera} {et~al.}(2011){Desidera}, {Carolo}, {Gratton}, {Martinez
  Fiorenzano}, {Endl}, {Mesa}, {Barbieri}, {Bonavita}, {Cecconi}, {Claudi},
  {Cosentino}, {Marzari}, \& {Scuderi}}]{desidera11}
{Desidera}, S., {Carolo}, E., {Gratton}, R., {et~al.} 2011, \aap, 533, A90

\bibitem[{{Doyle} {et~al.}(2011){Doyle}, {Carter}, {Fabrycky}, {Slawson},
  {Howell}, {Winn}, {Orosz}, {Prcaronsa}, {Welsh}, {Quinn}, {Latham}, {Torres},
  {Buchhave}, {Marcy}, {Fortney}, {Shporer}, {Ford}, {Lissauer}, {Ragozzine},
  {Rucker}, {Batalha}, {Jenkins}, {Borucki}, {Koch}, {Middour}, {Hall},
  {McCauliff}, {Fanelli}, {Quintana}, {Holman}, {Caldwell}, {Still},
  {Stefanik}, {Brown}, {Esquerdo}, {Tang}, {Furesz}, {Geary}, {Berlind},
  {Calkins}, {Short}, {Steffen}, {Sasselov}, {Dunham}, {Cochran}, {Boss},
  {Haas}, {Buzasi}, \& {Fischer}}]{doyle11}
{Doyle}, L.~R., {Carter}, J.~A., {Fabrycky}, D.~C., {et~al.} 2011, Science,
  333, 1602

\bibitem[{{Duch{\^e}ne}(2010)}]{duchene10}
{Duch{\^e}ne}, G. 2010, \apjl, 709, L114

\bibitem[{{Duch{\^e}ne} {et~al.}(2006){Duch{\^e}ne}, {Beust}, {Adjali},
  {Konopacky}, \& {Ghez}}]{duchene06}
{Duch{\^e}ne}, G., {Beust}, H., {Adjali}, F., {Konopacky}, Q.~M., \& {Ghez},
  A.~M. 2006, \aap, 457, L9

\bibitem[{{Duch{\^e}ne} {et~al.}(2007){Duch{\^e}ne}, {Delgado-Donate},
  {Haisch}, {Loinard}, \& {Rodr{\'{\i}}guez}}]{duchene07}
{Duch{\^e}ne}, G., {Delgado-Donate}, E., {Haisch}, Jr., K.~E., {Loinard}, L.,
  \& {Rodr{\'{\i}}guez}, L.~F. 2007, Protostars and Planets V, 379

\bibitem[{{Duch{\^e}ne} {et~al.}(2003){Duch{\^e}ne}, {Ghez}, {McCabe}, \&
  {Weinberger}}]{duchene03}
{Duch{\^e}ne}, G., {Ghez}, A.~M., {McCabe}, C., \& {Weinberger}, A.~J. 2003,
  \apj, 592, 288

\bibitem[{{Dupuy} \& {Liu}(2011)}]{dupuy11}
{Dupuy}, T.~J., \& {Liu}, M.~C. 2011, \apj, 733, 122

\bibitem[{{Duquennoy} \& {Mayor}(1991)}]{dm91}
{Duquennoy}, A., \& {Mayor}, M. 1991, \aap, 248, 485

\bibitem[{{Feigelson} \& {Nelson}(1985)}]{feigelson85}
{Feigelson}, E.~D., \& {Nelson}, P.~I. 1985, \apj, 293, 192

\bibitem[{{Fischer} \& {Marcy}(1992)}]{fm92}
{Fischer}, D.~A., \& {Marcy}, G.~W. 1992, \apj, 396, 178

\bibitem[{{Ghez} {et~al.}(1993){Ghez}, {Neugebauer}, \& {Matthews}}]{ghez93}
{Ghez}, A.~M., {Neugebauer}, G., \& {Matthews}, K. 1993, \aj, 106, 2005

\bibitem[{{Guilloteau} {et~al.}(1999){Guilloteau}, {Dutrey}, \&
  {Simon}}]{guilloteau99}
{Guilloteau}, S., {Dutrey}, A., \& {Simon}, M. 1999, \aap, 348, 570

\bibitem[{{Hartmann} {et~al.}(1998){Hartmann}, {Calvet}, {Gullbring}, \&
  {D'Alessio}}]{hartmann98}
{Hartmann}, L., {Calvet}, N., {Gullbring}, E., \& {D'Alessio}, P. 1998, \apj,
  495, 385

\bibitem[{{Ho} {et~al.}(2004){Ho}, {Moran}, \& {Lo}}]{ho04}
{Ho}, P.~T.~P., {Moran}, J.~M., \& {Lo}, K.~Y. 2004, \apjl, 616, L1

\bibitem[{{Hughes} {et~al.}(2008){Hughes}, {Wilner}, {Qi}, \&
  {Hogerheijde}}]{hughes08}
{Hughes}, A.~M., {Wilner}, D.~J., {Qi}, C., \& {Hogerheijde}, M.~R. 2008, \apj,
  678, 1119

\bibitem[{{Itoh} {et~al.}(2005){Itoh}, {Hayashi}, {Tamura}, {Tsuji}, {Oasa},
  {Fukagawa}, {Hayashi}, {Naoi}, {Ishii}, {Mayama}, {Morino}, {Yamashita},
  {Pyo}, {Nishikawa}, {Usuda}, {Murakawa}, {Suto}, {Oya}, {Takato}, {Ando},
  {Miyama}, {Kobayashi}, \& {Kaifu}}]{itoh05}
{Itoh}, Y., {Hayashi}, M., {Tamura}, M., {et~al.} 2005, \apj, 620, 984

\bibitem[{{Jensen} \& {Akeson}(2003)}]{jensen03}
{Jensen}, E.~L.~N., \& {Akeson}, R.~L. 2003, \apj, 584, 875

\bibitem[{{Jensen} {et~al.}(2007){Jensen}, {Dhital}, {Stassun}, {Patience},
  {Herbst}, {Walter}, {Simon}, \& {Basri}}]{jensen07}
{Jensen}, E.~L.~N., {Dhital}, S., {Stassun}, K.~G., {et~al.} 2007, \aj, 134,
  241

\bibitem[{{Jensen} {et~al.}(1996{\natexlab{a}}){Jensen}, {Koerner}, \&
  {Mathieu}}]{jensen96b}
{Jensen}, E.~L.~N., {Koerner}, D.~W., \& {Mathieu}, R.~D. 1996{\natexlab{a}},
  \aj, 111, 2431

\bibitem[{{Jensen} {et~al.}(2004){Jensen}, {Mathieu}, {Donar}, \&
  {Dullighan}}]{jensen04}
{Jensen}, E.~L.~N., {Mathieu}, R.~D., {Donar}, A.~X., \& {Dullighan}, A. 2004,
  \apj, 600, 789

\bibitem[{{Jensen} {et~al.}(1994){Jensen}, {Mathieu}, \& {Fuller}}]{jensen94}
{Jensen}, E.~L.~N., {Mathieu}, R.~D., \& {Fuller}, G.~A. 1994, \apjl, 429, L29

\bibitem[{{Jensen} {et~al.}(1996{\natexlab{b}}){Jensen}, {Mathieu}, \&
  {Fuller}}]{jensen96}
---. 1996{\natexlab{b}}, \apj, 458, 312

\bibitem[{{K{\"o}hler} {et~al.}(2008){K{\"o}hler}, {Ratzka}, {Herbst}, \&
  {Kasper}}]{kohler08}
{K{\"o}hler}, R., {Ratzka}, T., {Herbst}, T.~M., \& {Kasper}, M. 2008, \aap,
  482, 929

\bibitem[{{Kraus} \& {Hillenbrand}(2009{\natexlab{a}})}]{kraus09b}
{Kraus}, A.~L., \& {Hillenbrand}, L.~A. 2009{\natexlab{a}}, \apj, 704, 531

\bibitem[{{Kraus} \& {Hillenbrand}(2009{\natexlab{b}})}]{kraus09}
---. 2009{\natexlab{b}}, \apj, 703, 1511

\bibitem[{{Kraus} {et~al.}(2012){Kraus}, {Ireland}, {Hillenbrand}, \&
  {Martinache}}]{kraus11b}
{Kraus}, A.~L., {Ireland}, M.~J., {Hillenbrand}, L.~A., \& {Martinache}, F.
  2012, ArXiv e-prints

\bibitem[{{Kraus} {et~al.}(2011){Kraus}, {Ireland}, {Martinache}, \&
  {Hillenbrand}}]{kraus11}
{Kraus}, A.~L., {Ireland}, M.~J., {Martinache}, F., \& {Hillenbrand}, L.~A.
  2011, ArXiv e-prints

\bibitem[{{Lada}(2006)}]{lada06}
{Lada}, C.~J. 2006, \apjl, 640, L63

\bibitem[{{Launhardt}(2004)}]{launhardt04}
{Launhardt}, R. 2004, in IAU Symposium, Vol. 221, Star Formation at High
  Angular Resolution, ed. {M.~G.~Burton, R.~Jayawardhana, \& T.~L.~Bourke}, 213

\bibitem[{{Leinert} {et~al.}(1993){Leinert}, {Zinnecker}, {Weitzel},
  {Christou}, {Ridgway}, {Jameson}, {Haas}, \& {Lenzen}}]{leinert93}
{Leinert}, C., {Zinnecker}, H., {Weitzel}, N., {et~al.} 1993, \aap, 278, 129

\bibitem[{{Lin} \& {Papaloizou}(1979{\natexlab{a}})}]{lin79a}
{Lin}, D.~N.~C., \& {Papaloizou}, J. 1979{\natexlab{a}}, \mnras, 188, 191

\bibitem[{{Lin} \& {Papaloizou}(1979{\natexlab{b}})}]{lin79b}
---. 1979{\natexlab{b}}, \mnras, 186, 799

\bibitem[{{Loinard} {et~al.}(2007){Loinard}, {Torres}, {Mioduszewski},
  {Rodr{\'{\i}}guez}, {Gonz{\'a}lez-L{\'o}pezlira}, {Lachaume}, {V{\'a}zquez},
  \& {Gonz{\'a}lez}}]{loinard07}
{Loinard}, L., {Torres}, R.~M., {Mioduszewski}, A.~J., {et~al.} 2007, \apj,
  671, 546

\bibitem[{{Luhman} {et~al.}(2010){Luhman}, {Allen}, {Espaillat}, {Hartmann}, \&
  {Calvet}}]{luhman10}
{Luhman}, K.~L., {Allen}, P.~R., {Espaillat}, C., {Hartmann}, L., \& {Calvet},
  N. 2010, \apjs, 186, 111

\bibitem[{{Mannings} \& {Emerson}(1994)}]{mannings94}
{Mannings}, V., \& {Emerson}, J.~P. 1994, \mnras, 267, 361

\bibitem[{{Markwardt}(2009)}]{markwardt09}
{Markwardt}, C.~B. 2009, in Astronomical Society of the Pacific Conference
  Series, Vol. 411, Astronomical Data Analysis Software and Systems XVIII, ed.
  {D.~A.~Bohlender, D.~Durand, \& P.~Dowler}, 251

\bibitem[{{Massarotti} {et~al.}(2005){Massarotti}, {Latham}, {Torres}, {Brown},
  \& {Oppenheimer}}]{massarotti05}
{Massarotti}, A., {Latham}, D.~W., {Torres}, G., {Brown}, R.~A., \&
  {Oppenheimer}, B.~D. 2005, \aj, 129, 2294

\bibitem[{{Massi} {et~al.}(2006){Massi}, {Forbrich}, {Menten},
  {Torricelli-Ciamponi}, {Neidh{\"o}fer}, {Leurini}, \& {Bertoldi}}]{massi06}
{Massi}, M., {Forbrich}, J., {Menten}, K.~M., {et~al.} 2006, \aap, 453, 959

\bibitem[{{Mathieu}(1994)}]{mathieu94}
{Mathieu}, R.~D. 1994, \araa, 32, 465

\bibitem[{{Mathieu} {et~al.}(2000){Mathieu}, {Ghez}, {Jensen}, \&
  {Simon}}]{mathieu00}
{Mathieu}, R.~D., {Ghez}, A.~M., {Jensen}, E.~L.~N., \& {Simon}, M. 2000,
  Protostars and Planets IV, 703

\bibitem[{{Mathieu} {et~al.}(1997){Mathieu}, {Stassun}, {Basri}, {Jensen},
  {Johns-Krull}, {Valenti}, \& {Hartmann}}]{mathieu97}
{Mathieu}, R.~D., {Stassun}, K., {Basri}, G., {et~al.} 1997, \aj, 113, 1841

\bibitem[{{McCabe} {et~al.}(2011){McCabe}, {Duch{\^e}ne}, {Pinte},
  {Stapelfeldt}, {Ghez}, \& {M{\'e}nard}}]{mccabe11}
{McCabe}, C., {Duch{\^e}ne}, G., {Pinte}, C., {et~al.} 2011, \apj, 727, 90

\bibitem[{{Melo} {et~al.}(2001){Melo}, {Covino}, {Alcal{\'a}}, \&
  {Torres}}]{melo01}
{Melo}, C.~H.~F., {Covino}, E., {Alcal{\'a}}, J.~M., \& {Torres}, G. 2001,
  \aap, 378, 898

\bibitem[{{Moriarty-Schieven} {et~al.}(2006){Moriarty-Schieven}, {Johnstone},
  {Bally}, \& {Jenness}}]{moriarty-schieven06}
{Moriarty-Schieven}, G.~H., {Johnstone}, D., {Bally}, J., \& {Jenness}, T.
  2006, \apj, 645, 357

\bibitem[{{Motte} \& {Andr{\'e}}(2001)}]{motte01}
{Motte}, F., \& {Andr{\'e}}, P. 2001, \aap, 365, 440

\bibitem[{{Mugrauer} \& {Neuh{\"a}user}(2009)}]{mugrauer09}
{Mugrauer}, M., \& {Neuh{\"a}user}, R. 2009, \aap, 494, 373

\bibitem[{{Mundy} {et~al.}(1996){Mundy}, {Looney}, {Erickson}, {Grossman},
  {Welch}, {Forster}, {Wright}, {Plambeck}, {Lugten}, \& {Thornton}}]{mundy96}
{Mundy}, L.~G., {Looney}, L.~W., {Erickson}, W., {et~al.} 1996, \apjl, 464,
  L169

\bibitem[{{Nagel} {et~al.}(2010){Nagel}, {D'Alessio}, {Calvet}, {Espaillat},
  {Sargent}, {Hern{\'a}ndez}, \& {Forrest}}]{nagel10}
{Nagel}, E., {D'Alessio}, P., {Calvet}, N., {et~al.} 2010, \apj, 708, 38

\bibitem[{{Ochi} {et~al.}(2005){Ochi}, {Sugimoto}, \& {Hanawa}}]{ochi05}
{Ochi}, Y., {Sugimoto}, K., \& {Hanawa}, T. 2005, \apj, 623, 922

\bibitem[{{Osterloh} \& {Beckwith}(1995)}]{osterloh95}
{Osterloh}, M., \& {Beckwith}, S.~V.~W. 1995, \apj, 439, 288

\bibitem[{{Pani{\'c}} {et~al.}(2009){Pani{\'c}}, {Hogerheijde}, {Wilner}, \&
  {Qi}}]{panic09}
{Pani{\'c}}, O., {Hogerheijde}, M.~R., {Wilner}, D., \& {Qi}, C. 2009, \aap,
  501, 269

\bibitem[{{Papaloizou} \& {Pringle}(1977)}]{papaloizou77}
{Papaloizou}, J., \& {Pringle}, J.~E. 1977, \mnras, 181, 441

\bibitem[{{Patience} {et~al.}(2008){Patience}, {Akeson}, \&
  {Jensen}}]{patience08}
{Patience}, J., {Akeson}, R.~L., \& {Jensen}, E.~L.~N. 2008, \apj, 677, 616

\bibitem[{{Patience} {et~al.}(2002){Patience}, {White}, {Ghez}, {McCabe},
  {McLean}, {Larkin}, {Prato}, {Kim}, {Lloyd}, {Liu}, {Graham}, {Macintosh},
  {Gavel}, {Max}, {Bauman}, {Olivier}, {Wizinowich}, \& {Acton}}]{patience02}
{Patience}, J., {White}, R.~J., {Ghez}, A.~M., {et~al.} 2002, \apj, 581, 654

\bibitem[{{Pichardo} {et~al.}(2005){Pichardo}, {Sparke}, \&
  {Aguilar}}]{pichardo05}
{Pichardo}, B., {Sparke}, L.~S., \& {Aguilar}, L.~A. 2005, \mnras, 359, 521

\bibitem[{{Pi{\'e}tu} {et~al.}(2011){Pi{\'e}tu}, {Gueth}, {Hily-Blant},
  {Schuster}, \& {Pety}}]{pietu11}
{Pi{\'e}tu}, V., {Gueth}, F., {Hily-Blant}, P., {Schuster}, K.-F., \& {Pety},
  J. 2011, \aap, 528, A81

\bibitem[{{Raghavan} {et~al.}(2006){Raghavan}, {Henry}, {Mason}, {Subasavage},
  {Jao}, {Beaulieu}, \& {Hambly}}]{raghavan06}
{Raghavan}, D., {Henry}, T.~J., {Mason}, B.~D., {et~al.} 2006, \apj, 646, 523

\bibitem[{{Raghavan} {et~al.}(2010){Raghavan}, {McAlister}, {Henry}, {Latham},
  {Marcy}, {Mason}, {Gies}, {White}, \& {ten Brummelaar}}]{raghavan10}
{Raghavan}, D., {McAlister}, H.~A., {Henry}, T.~J., {et~al.} 2010, \apjs, 190,
  1

\bibitem[{{Reipurth} \& {Zinnecker}(1993)}]{rz93}
{Reipurth}, B., \& {Zinnecker}, H. 1993, \aap, 278, 81

\bibitem[{{Richichi} {et~al.}(1999){Richichi}, {K{\"o}hler}, {Woitas}, \&
  {Leinert}}]{richichi99}
{Richichi}, A., {K{\"o}hler}, R., {Woitas}, J., \& {Leinert}, C. 1999, \aap,
  346, 501

\bibitem[{{Schaefer} {et~al.}(2009){Schaefer}, {Dutrey}, {Guilloteau}, {Simon},
  \& {White}}]{schaefer09}
{Schaefer}, G.~H., {Dutrey}, A., {Guilloteau}, S., {Simon}, M., \& {White},
  R.~J. 2009, \apj, 701, 698

\bibitem[{{Schaefer} {et~al.}(2006){Schaefer}, {Simon}, {Beck}, {Nelan}, \&
  {Prato}}]{schaefer06}
{Schaefer}, G.~H., {Simon}, M., {Beck}, T.~L., {Nelan}, E., \& {Prato}, L.
  2006, \aj, 132, 2618

\bibitem[{{Scholz} {et~al.}(2006){Scholz}, {Jayawardhana}, \&
  {Wood}}]{scholz06}
{Scholz}, A., {Jayawardhana}, R., \& {Wood}, K. 2006, \apj, 645, 1498

\bibitem[{{Simon} {et~al.}(1992){Simon}, {Chen}, {Howell}, {Benson}, \&
  {Slowik}}]{simon92}
{Simon}, M., {Chen}, W.~P., {Howell}, R.~R., {Benson}, J.~A., \& {Slowik}, D.
  1992, \apj, 384, 212

\bibitem[{{Simon} {et~al.}(2000){Simon}, {Dutrey}, \& {Guilloteau}}]{simon00}
{Simon}, M., {Dutrey}, A., \& {Guilloteau}, S. 2000, \apj, 545, 1034

\bibitem[{{Simon} {et~al.}(1995){Simon}, {Ghez}, {Leinert}, {Cassar}, {Chen},
  {Howell}, {Jameson}, {Matthews}, {Neugebauer}, \& {Richichi}}]{simon95}
{Simon}, M., {Ghez}, A.~M., {Leinert}, C., {et~al.} 1995, \apj, 443, 625

\bibitem[{{Tokovinin}(1998)}]{tokovinin98}
{Tokovinin}, A.~A. 1998, Astronomy Letters, 24, 178

\bibitem[{{Torres}(1999)}]{torres99}
{Torres}, G. 1999, \pasp, 111, 169

\bibitem[{{Torres} {et~al.}(2007){Torres}, {Loinard}, {Mioduszewski}, \&
  {Rodr{\'{\i}}guez}}]{torres07}
{Torres}, R.~M., {Loinard}, L., {Mioduszewski}, A.~J., \& {Rodr{\'{\i}}guez},
  L.~F. 2007, \apj, 671, 1813

\bibitem[{{Torres} {et~al.}(2009){Torres}, {Loinard}, {Mioduszewski}, \&
  {Rodr{\'{\i}}guez}}]{torres09}
---. 2009, \apj, 698, 242

\bibitem[{{Verrier} \& {Evans}(2008)}]{verrier08}
{Verrier}, P.~E., \& {Evans}, N.~W. 2008, \mnras, 390, 1377

\bibitem[{{White} \& {Ghez}(2001)}]{wg01}
{White}, R.~J., \& {Ghez}, A.~M. 2001, \apj, 556, 265

\bibitem[{{White} {et~al.}(1999){White}, {Ghez}, {Reid}, \&
  {Schultz}}]{white99}
{White}, R.~J., {Ghez}, A.~M., {Reid}, I.~N., \& {Schultz}, G. 1999, \apj, 520,
  811

\bibitem[{{White} \& {Hillenbrand}(2005)}]{white05}
{White}, R.~J., \& {Hillenbrand}, L.~A. 2005, \apjl, 621, L65

\bibitem[{{Zahn}(1977)}]{zahn77}
{Zahn}, J.-P. 1977, \aap, 57, 383

\bibitem[{{Zahn} \& {Bouchet}(1989)}]{zahn89}
{Zahn}, J.-P., \& {Bouchet}, L. 1989, \aap, 223, 112

\end{thebibliography}
